\documentclass[a4, 12pt]{article}
    \usepackage{lmodern}
    \usepackage[round]{natbib}
    \usepackage{longtable}
    \usepackage{geometry}
    \usepackage{array}
    \usepackage{multirow}
    \usepackage{wrapfig}
    \usepackage{pdflscape}
    \usepackage{tabu}
    \usepackage[normalem]{ulem}
    \usepackage{makecell}

    \usepackage{fullpage}
    \usepackage{amssymb}
    \usepackage{float}
    \usepackage{amsmath}
    \usepackage{bbm}
    \usepackage{mathtools}
    \usepackage{dsfont}
    \usepackage{enumerate}
    \usepackage{bm}
    \usepackage[Export]{adjustbox}
    
    \usepackage{booktabs}
    \usepackage{dcolumn} 
    \newcolumntype{d}[1]{D{.}{.}{#1}}
    \usepackage[flushleft]{threeparttable} 
    \usepackage{threeparttablex}
    \usepackage{tabularx}
        \usepackage{cellspace}
    \usepackage[font=sc, labelsep=period]{caption}
    \usepackage{chngcntr}
    \usepackage{etoolbox}
    \pretocmd{\appendix}{%
    \counterwithin*{figure}{section}%
    \counterwithin*{table}{section}%
    \setcounter{figure}{0}\setcounter{table}{0}%
    }{}{}

    \usepackage{placeins} 

    \usepackage{subcaption}

    \usepackage{graphicx}
    \graphicspath{{./figures/}}

    \usepackage[pagebackref=true]{hyperref}
    \renewcommand*{\backref}[1]{}
    \renewcommand*{\backrefalt}[4]{%
        \ifcase #1 (Not cited.)%
        \or        [Cited on page~#2.]%
        \else      [Cited on pages~#2.]%
        \fi}
    \usepackage{url}
        
    \usepackage[utf8]{inputenc}

    \usepackage{scalerel,stackengine}
    \stackMath
    \newcommand\reallywidehat[1]{%
    \savestack{\tmpbox}{\stretchto{%
    \scaleto{%
        \scalerel*[\widthof{\ensuremath{#1}}]{\kern.1pt\mathchar"0362\kern.1pt}%
        {\rule{0ex}{\textheight}}
    }{\textheight}%
    }{2.4ex}}%
    \stackon[-6.9pt]{#1}{\tmpbox}%
    }
   \usepackage{titlesec}
    \titleformat*{\section}{\large\bfseries}
    \titleformat*{\subsection}{\normalsize\bfseries}
    \titleformat*{\subsubsection}{\normalsize\itshape}
    \titleformat*{\paragraph}{\normalsize\itshape}

    \usepackage{wrapfig}  
    \usepackage{verbatim}  
    \usepackage{lscape}  

    \usepackage[table,xcdraw]{xcolor}

    \usepackage{tikz}
        \usetikzlibrary{calc}
        \usetikzlibrary{trees}
        \usetikzlibrary{decorations.pathreplacing,angles,quotes}
        \usetikzlibrary{positioning, matrix}

    \usepackage{titling}
    \makeatletter
    \def\input@path{{./tables/}}
    \makeatother

    \usepackage{setspace}
\begin{document}

\begin{singlespace}
    \title{Legal aid eligibility and court outcomes:
        a design-based double-machine-learning approach}

    \author{Fabio I. Martinenghi%
        \thanks{School of Economics, University of New South Wales, and
        Newcastle Business School, University of Newcastle.
        Email: fabio.martinenghi@newcastle.edu.au.
        I am grateful to Don Weatherburn for his mentorship and for
        facilitating contact with Legal Aid NSW. I thank Legal Aid NSW
        for giving me access to their dataset and for their outstanding
        support. The views expressed in this paper are my own and not
        necessarily those of Legal Aid NSW. I thank Richard Holden for
        his financial support, Fangzhou Yu and Tim Neal for their helpful
        comments, and all the participants at seminars and conferences
        at Bocconi University, the University of Melbourne, the
        University of Technology Sydney, and the Australian National
        University for their feedback.
        }}
    \date{}
    \maketitle

    \begin{abstract}
        Equality before the law is a human right, and access to
        high-quality legal aid for indigent defendants is essential to
        enforce it.
        In a context where all defendants have access to a lawyer,
        I study the impact of denying legal aid on court outcomes.
        I combine double machine learning and a new administrative dataset
        linking aid applications to court outcomes
        in New South Wales, Australia, to learn the
        assignment function, whose inputs are known.
        I find that applicants who fail the means test and
        hire private lawyers are 10
        percentage points less likely to be incarcerated than if they passed
        and relied on legal aid. Given an average incarceration length of nearly
        four years, this gap is significant.
        However, I find evidence suggesting that they spend more time
        in jail if they are incarcerated.
        A government preference for broad access to aid over allocated
        time per case could explain this pattern.
    \end{abstract}
\end{singlespace}

\section{Introduction}
The principle of equality before the law states that everyone's rights should
be equally protected by the law without discrimination.
It is an important and ancient democratic principle,
dating at least back to the
fifth century BC, when it was mentioned by \cite{thucydides431}
in praise of Athenian institutions.
More recently, the United Nations General Assembly (\citeyear{UDHR})
recognised it as universal human right .
One way in which governments implement this principle is by offering
``legal advice, assistance and representation […]
at no cost for those without sufficient means''\footnote{
    This is the definition of the right to counsel, which the
    government enforces via legal aid.
} \citep{un2012}.
This is also called ``legal aid''.
While legal aid is necessary to achieve some equality before the law,
the extent to which this equality is realised depends on the quality of the
services that constitute the legal aid, that is, the performance of publicly-funded
compared to privately-funded legal representation.
Intuitively, the more accessing privately-funded lawyers improves a
defendant's outcomes in court compared to a publicly-funded lawyer,
the less the justice system has achieved
equality before the law.

In this paper, I study the effect of denying legal aid on applicants' case
outcomes. This is a first in the literature.
I use administrative data on legal aid applicants charged with serious crimes
in New South Wales (NSW), Australia.
For such cases, legal aid can only be denied if the applicant fails a means
test\textemdash that is, if they are found capable of affording private
legal representation.
Hence, by construction, (i) approximately 98\% of legal aid applicants are
legally represented and only 2\% self-represent\textemdash
in this study, we focus on the former\textemdash and (ii)
applicants are not assigned aid at random, so that
estimated treatment effects that rely on mean comparisons between
successful and unsuccessful applicants are marred by selection bias and
fail identification.
In other words, only defendants with high levels of disadvantage are
granted legal aid, and such disadvantage may influence their chances
of being convicted 
(for example, see \citeauthor{denadai2020}, \citeyear{denadai2020}, or,
in Australia, \citeauthor{don1998}, \citeyear{don1998}).

Moreover, although the means tests set explicit thresholds for income and
assets, eligibility is ultimately determined after combining these
\textit{rules} with \textit{standards}.
First, as part of its Means Test, Legal Aid NSW conducts an
ability-to-pay assessment that can deny aid when the applicant is
judged able to finance private representation, based on their borrowing capacity
and whether their lifestyle indicates access to funds beyond those
declared \citep{laPolicies}.
Second, Legal Aid NSW may grant aid to applicants who fail the threshold
tests where exceptional circumstances apply,
based on factors such as the likely cost of the proceedings, and
the applicant's overall financial position \citep{laPolicies}.
These standards introduce an element of bounded discretion into what would
otherwise be a mechanical rule
\citep[see][for a classic discussion of rules versus standards]{sullivan1992}.
As a result, while the dataset includes the inputs used in aid decisions,
the analyst does not observe the functional form of the treatment-assignment
mechanism\textemdash how those inputs map into an approval or
denial decision, or which dimensions receive greater weight.
In short, the assignment rule is latent\textemdash
it is not given ex ante and must be learned from the data.

To address this econometric challenge, I take a double machine learning approach
\citep[DML; ][]{chern2018}.
Specifically, I use the random forests algorithm \citep{breiman2001}
on administrative data that include all the inputs of the treatment
assignment function\textemdash that is, all the 70+ factors derived
from the legal aid application form and considered
when deciding whether to grant aid.
This allows me to flexibly learn the treatment assignment function and
use the resulting propensity score (jointly with a flexible outcome regression)
to recover treatment effects under uncounfoundedness \citep[][]{chern2018}.
Indeed, as highlighted in \cite{wager2025}, double machine learning
estimators are a type of augmented inverse probability weighting estimator
\citep[AIPW; first proposed in][]{robins1994}, and as such they are consistent
and efficient under unconfoundedness.

I find that aid denial
decreases the probability of a defendant going to jail
by 8.1 percentage points (p.p.),
compared to being represented by a public lawyer.
This effect is sizeable, particularly considering that positive
jail sentences average 46.5 months\textemdash almost four years.
Moreover, aid denial does not significantly reduce incarceration time,
while, for those sentenced to jail time, aid denial seems to increase
incarceration time.
Finally, aid denial decreases
the chances of pleading guilty to the highest charge by 4.6 p.p. and
increases the chances of being fined by 4.8 p.p.
I interpret these findings as evidence that public lawyers rely more
heavily on plea bargaining than private lawyers\textemdash a time-efficient
strategy that is more likely to lead to incarceration but on average lowers the
incarceration spell. This pattern of behaviour is possibly due to their limited
budget and high workload, combined with a prioritisation of access to justice
over time per case.

In a separate analysis, I further investigate the performance gap between the
private and public legal services by dropping from my sample applicants
who self-represented.
I find that failing to receive aid and hiring of a private lawyer
decreases the defendant's probability of incarceration by 9.7 p.p.,
seems to lengthen the incarceration spell for those who are sentenced
to jail,
decreases the chances of pleading guilty to the highest charge by 5 p.p.
and increases those of being fined by 6.6 p.p.
Furthermore, the privately funded lawyers observed in my sample are plausibly
drawn from the lower-fee segment of the private market, since they are retained
by individuals whose income is low enough to \textit{potentially} qualify for
legal aid. If lawyer quality is increasing in fees, then my estimated
performance difference between publicly funded and
privately funded representation is conservative\textemdash
that is, a lower bound on the performance difference of Legal Aid against the
\textit{full} private market, including higher-fee lawyers.

I rely on two administrative datasets.
First, the Legal Aid NSW dataset covers all applications for aid made between
2012 and 2021 and related to serious criminal cases.
Second, each application is linked to its court data, stored in
the Re-Offending Database, which is curated by the
NSW Bureau of Crime Statistics and Research (BOCSAR).
This linkage provides me with a wealth of information on the applicant
(for example, country of origin, number of dependants, year of birth), the
application (for example, time of submission, which offices processed it) and the
case (for example, trial length, sentence).

The contribution of this paper is two-fold, as it both adds to the indigent
legal defence literature and the causal machine learning literature.
On the indigent defence front
\citep[see][]{anderson2012,iyengar2007,roach2014,shemtov2022,agan2021,schwall2017,lee2021},
it is the first study to provide evidence on
the effect of legal aid on court outcomes.
The existing empirical research around indigent defence
does not address this point, as it mostly leverages
institutional designs and reforms in the public defence context
to study how lawyers respond to different
incentive schemes. One strand of the literature studies
how the court outcomes of indigent defendants
change when legal aid comes in the form of
publicly-funded private lawyers (called ``panel lawyers'' in Australia, and
``assigned counsels'' in the US)
compared to public lawyers
\citep[``in-house lawyers'' in Australia, ``public defenders'' in the US; see][]{
    anderson2012,iyengar2007,roach2014,shemtov2022}.
This literature consistently shows that panel lawyers tend to perform worse than
public lawyers. Because panel lawyers are simply private lawyers who have taken
a publicly-funded case, these results could suggest that
counterfactual private lawyers perform worse than public lawyers\textemdash
although this depends on how panel lawyers are remunerated.

Relatedly, \cite{porreca2023} study the introduction of the right to
counsel from a historical perspective, finding difference-in-differences
evidence that it led to an increase in conviction rates in 19th century London.
Their study, while related to mine, speaks more to the issue of the
professionalisation of criminal defense than that of legal aid provision\textemdash 
as they note, a publicly-funded legal aid society was only
established in 1949.

Another strand of the literature
further investigates how fee structure affects
legal performance. They do this by focusing on reforms that change
the compensation structure of a single type of lawyer, panel lawyers
\citep{schwall2017,lee2021}.
\cite{schwall2017} and \cite{lee2021} both use a difference-in-differences
design and find that, compared to hourly rates, flat fees drastically reduce
the number of hours worked by defence lawyers while either having no detectable
impact on court outcomes \citep{schwall2017} or a large negative impact on
the same \citep{lee2021}.
In-house lawyers work under a fixed salary, as opposed to the private and
panel lawyers' hourly rates. Hence, this strand of the literature would suggest
that in-house lawyers perform worse than private lawyers.
It is, therefore, ex-ante unclear whether in-house lawyers
perform better or worse than private lawyers\textemdash an open
empirical question that this paper sets to investigate.

Furthermore, this study contributes to the causal machine learning literature
by providing a first application on double machine learning
design-based identification.
It also illustrates how to increase the transparency of DML
by visualising its estimated propensity score.
While the methodological literature on causal machine learning is
mature, applications remain scant within Economics.
In randomised control trials (RCT) settings,
\cite{bertrand2021} and \cite{davis2020} leverage
causal random forests \citep{wager2018} to recover individual treatment effects.
\cite{chern2018generic} develop and apply an ML method to study treatment
effect heterogeneity in an RCT context.
Similarly, \cite{knaus2022b} develop and apply a procedure that combines ML and
impact evaluation techniques to heterogeneity in RCTs.
In observational settings, the main published papers using ML for estimating
treatment effects are \cite{deryugina2019}, who use random forests and LASSO
to predict the life-years lost due to air pollution and apply
\citep{chern2018generic} to study treatment effect heterogeneity. Finally,
\cite{knaus2022a} applies DML to longitudinal survey data
to estimate the (dose-response) effects of musical practice on students'
cognitive skills.

In particular, this application shows how the identification strategy relying on
conditional independence (or \textit{unconfoundedness}), generally considered
weak in the economic literature, can be credible in the right context.
This is the case in observational impact evaluation settings where
the variables used for assigning treatment are known and observed, but
functional form of the assignment mechanism is not known. Indeed, while some
policies make use of sharp rules for assigning treatment, many rely
on fuzzier rules that leave space for discretion, or may be based on sharp rules
in theory but not in practice. In such cases, relying
on more standard methods such as regression discontinuity
designs not only restricts the
target parameter to a local average treatment effect, but it
further focuses on the \textit{complier} fraction of such population, leading
to potential sample-size challenges.
Instead, using DML with flexible algorithms such as random forests allows
to credibly estimate the treatment assignment function while also
targeting global parameters like the average treatment effect (ATE)
and the average treatment effect on the
treated (ATT), serving as either a complement or a substitute
to standard estimators.

Moreover, a concern about the transparency of machine learning methods seems
to be a barrier to their widespread adoption in economics research.
While more familiar methods, such as difference in differences or
regression discontinuity design, are associated with
a wide array of tests to assess their robustness and the strength
of their identifying assumptions,
analogous tests are less present and known in the ML literature.
To address this issue, I show that the propensity score estimated by the
DML Interactive Regression Model in \cite{chern2018}
can be used to assess the credibility of the common support assumption
and the sensitivity of the results to including observations with low overlap.
This is in a similar spirit to \cite{knaus2022b},
which adapts a covariate balance test
to DML.

From a policy implication perspective, the approach
taken in this paper can be applied to monitor potential performance gaps between
public and private legal services, and extended to other policy evaluation
exercises.
Moreover, the very ability to measure this gap opens new policy questions,
allowing citizens to form an opinion over whether such a gap should exist, and
potentially advocate for policies to reduce it.
Because publicly-funded lawyers are typically paid less than
private sector lawyers \citep[see][for NSW evidence]{forell2010},
a simple policy to reduce the gap could be to increase their
salary/rates. Indeed, the most recent available data show that
public and private lawyers earned, respectively, A\$86,700 and A\$94,800
on average in 2008/09 \citep{forell2010},
while public and private barristers earned,
respectively, A\$106,184 \citep{forell2010} and A\$361,850
\citep{abs_legal_2008}\footnote{
    Estimate calculated from ABS data dividing the income generated by
    private barristers in the 2007-08 financial year (A\$1.4b)
    by their number in the same period (3,869 private barristers).
    Public barristers are to be intended as senior public lawyers acting as
    barristers.
}
on average.

The remainder of the paper is organised as follows. Section \ref{sec:background}
presents the policy, its eligibility rules, the organisation providing aid and
the rules for its provision and some institutional context.
Section \ref{sec:data} introduces the dataset, while Section
\ref{sec:descriptive} presents some descriptive evidence from it.
Section \ref{sec:methods} presents the empirical approach taken to
estimate the treatment effects.
Section \ref{sec:results} presents the results and related sensitivity
analyses, Section \ref{sec:discuss}
discusses them, and
Section \ref{sec:conclusion} concludes.

\section{Context and Background \label{sec:background}}
\subsection{Legal Aid}
In Australia,
\cite{dietrich1992} establishes the right to a fair trail for federal cases,
which include those involving serious crimes (``indictable crimes'' in
Australia, ``felonies'' in the US).
In New South Wales, this has been interpreted as implying
the right to counsel\textemdash
the right to ``legal advice, assistance and representation
    […] at no cost for those without sufficient means'' \citep{un2012}\textemdash
so that defendants accused of serious crimes are offered the chance to apply for
legal aid.

Legal Aid NSW is a government organisation providing legal assistance to
disadvantaged people in New South Wales (NSW), Australia. It was
founded in 1979 under the name of Legal Services Commission, and it now counts
more than 1,600 staff and 25 offices, two satellite offices and 243
regular outreach locations spread across NSW \citep{laWhoWeAre}.

Legal Aid provides such legal services either directly,
via in-house lawyers and barristers,
or by funding private lawyers and barristers at an agreed
lower-than-market rate. Private lawyers working on Legal Aid matters are called
``panel lawyers'' (``assigned counsels''
in the US), a useful term that avoids confusing them with
private lawyers working on privately-funded matters.
Legal Aid covers (with restrictions) matters in Family, Civil and Criminal 
Law\textemdash both at a state and at a federal (or ``Commonwealth'') level.
In this paper, I focus on the Criminal Law Division of Legal Aid NSW, as it
is the one covering Criminal matters.

\subsection{Eligibility \label{sec:eligibility}}
Legal aid determines whether an applicant is eligible for aid based on four
criteria: the Jurisdiction test, the Means test, the Merit test and the
Availability of Funds test \citep{laBrochure}. Matters related to
indictable crimes automatically pass the Jurisdiction and the Merit tests
\textemdash the former with some rare exceptions detailed in Appendix
\ref{appx:tests}.
Historically, the availability of funds test has never been used.
It follows that the Means test is the only test relevant to our context.
Appendix \ref{appx:tests} includes a brief review of the other
Legal Aid tests for completeness.

The Means test focuses on the applicant's income and assets.
It seeks to
determine whether they are eligible for aid and how much they would need to
contribute towards their legal expenses if their application were successful.
Applicants are required to provide documentary evidence to verify their
declared income and assets\textemdash failing to do so results in an application
refusal. Moreover, knowingly making materially false statements in the
application is an offence that carries a maximum penalty of 5,500\$
(50 penalty units) or imprisonment for 6 months.
The documentary evidence typically includes (where applicable)
a recent payslip, a letter from the employer,
the most recent tax return, balance sheets, and bank statements showing
three months of operation on all bank accounts\textemdash and Legal Aid NSW
reserves the right to request further information.

The Means test consists of three sub-tests\textemdash income, assets and ability-to-pay
legal costs tests. To pass the income test, an applicant must earn less
than \$400 in \textit{net assessable income}\footnote{
    In Appendix \ref{appx:frd}, I attempt to use a fuzzy regression discontinuity
    design on this threshold, but am unable to produce credible treatment estimates
    due to several issues, including evidence of sorting around the cutoff.
}.
This is calculated as
gross income minus
some benefits and deductions, most of which are a function of the number of
the applicant's dependants
\citep[see][for details]{laPolicies}.
An implication of this test is that if an applicant is receiving the
maximum amount of welfare transfers, they pass the test.
The asset test sets the eligibility threshold at 100\$ in net assets, after
taking an applicant's (gross assessable) assets and deducting, subject to caps,
a range of items\textemdash such as home and business equity,
motor vehicle equity and furniture value.
Then, assets and income are considered together\textemdash jointly
with expenditures\textemdash to assess whether the
applicant can afford to hire private legal representatives, most likely
by securing a loan.
This third stage is the ability to pay legal costs test, which introduces a
standards-based assessment to complement the sharp income and asset rules.
Even when the income and assets tests are satisfied, an application may be
refused if Legal Aid NSW determines that the applicant can afford to pay,
based on (i) the applicant's general assets and ability to
realise them or secure a loan against their assets
(for example, home equity and business assets), and (ii) the applicant's lifestyle,
activities, and general expenditure \citep{laPolicies}.
It should be noted that the ability-to-pay test, although more discretionary than the
income and asset rules, is nevertheless evaluated using applicant information
that is recorded in our dataset (and available to Legal Aid officers).
Finally, every application passes the Availability of Funds test as long as
Legal Aid has the funding to provide its services. Historically,
this has always been the case.

\subsection{Serious Crimes}
I focus on serious crimes, which are called \textit{indictable offences}
in Australia (and \textit{felonies} in the US),
and which hence exclude the less serious
\textit{summary offences} (\textit{misdemeanors} in the US).
In NSW, criminal offences
can be catalogued as either Summary, Table 1, Table 2 and Strictly Indictable.
Table 1 and 2 offences can be deemed as either summary\textemdash and hence
processed by a magistrate in the Local Court\textemdash or indictable\textemdash
hence dealt with by a judge in the District or Supreme Court.
Instead, Summary Offences are always summary and Strictly Indictable are always
indictable.
Within indictable matters, legal aid is available for committals,
bail applications, mentions and adjournments, sentence matters,
variations or revocations of a
Community Correction Order or Conditional Release Order and criminal trials
\citep{laBrochure}.

\subsection{Daily Operations}

A defendant can apply for legal aid
by filling in a form and either posting, emailing, or hand-delivering
it to a Legal Aid office. Alternatively, they can have an in-house Legal Aid
lawyer or a private lawyer submit it for them\textemdash the former free of
charge, the latter according to their rate
(although some do not charge a fee).

However, the location where the application is lodged does not determine who
will handle the case. A case is assigned based on the court jurisdiction
under which it fell. Each Legal Aid office is responsible for at least one
court. Not all courts are covered by Legal Aid offices though,
and Legal Aid has to rely entirely on panel lawyers when a case falls into
a non-covered court.
Cases are also assigned to panel lawyers whenever the relevant Legal Aid
office has a conflict of interest or
if it is at capacity.
Moreover, whenever a case is assigned to a public (in-house) lawyer, the lawyer
is allocated on an as-good-as-random basis. In the present context,
cases received by a Legal NSW Crime Division office will be allocated to lawyers
within that office in no particular order.

Finally, an application for legal aid can either result in aid being granted,
denied (``refused''), or initially granted and then denied (``terminated'').
Applicants who are successful but immediately change their mind and opt for
a (privately-funded) private lawyer or self-representation
are also labelled as ``terminated''.
Applicants do not have a right of appeal against a refusal of aid made
``wholly or partly on the ground the applicant fails to satisfy
the means test'' \citep[][ss56 1AA]{LAact}\textemdash
meaning that applicants in this study's sample could
not appeal.

\subsection{Exceptions and the Role of Discretion}

As noted above, Legal Aid NSW does not determine eligibility exclusively from
the income and assets thresholds. Rather, the threshold criteria are
supplemented by standards-based assessments that can them
in either direction: aid may be denied even when the applicant meets
the income and assets
tests\textemdash via the ability-to-pay assessment\textemdash
and it may be granted even when the thresholds are
exceeded\textemdash via \textit{discretion}, presented below.
This is where bounded discretion enters the treatment
assignment mechanism.

Based on the ability-to-pay assessment,
a defendant passing the income and assets tests might be denied aid if their
lifestyle, as evidenced by the information included in their aid application
form, suggests that they have access to funds other than those declared.
For instance, a defendant might have an unstated informal income stream, or
have relatives abroad regularly transferring money to support a lifestyle
otherwise above their means. This could be reflected by their
expenditures being disproportionately higher than their income.

Moreover, a defendant failing these tests might be provided aid via
what is called \textit{discretion}, as stated in \cite{LAact} ss 30.
In practice, if an applicant fails the Means
test, the lawyer who was assigned the case can ask for discretion if the
applicant's financial position meets certain standards.
This request is then reviewed, and mostly granted, by
another member of staff.
In particular, aid can be provided via discretion depending on:
(i) the likely cost of the proceedings;
(ii) the type of proceedings;
(iii) the overall financial position of the applicant;
(iv) whether the applicant would suffer undue financial hardship if
legal aid was refused;
(v) whether the defendant has enough time to raise the required funds;
(vi) whether it is reasonable to expect the defendant to borrow against their
assets.

In sum, while the income and asset cutoffs seem to define an application system
built around sharp rules, the use of standards via discretion policies
imply a softer approach.
They define a system where Legal Aid staff is tasked with reviewing
the overall financial position of the applicant and use this information
to assess their ability to pay for private representation.


\section{Data \label{sec:data}}
The data used in this study covers the universe of New South Wales
criminal cases where the defendant was charged with a serious
(\textit{indictable}) crime, applied for legal aid
between 1 January 2012 and 31 December 2021, and is not recorded as
Aboriginal nor Torres Strait Islander\footnote{
    Access to Aboriginal and Torres Strait Islander data has additional
    ethics requirements that I could not fulfil within my timeline.
}. This is obtained by merging two administrative datasets.

The first dataset was provided by Legal Aid NSW and includes information on the
applicants to legal aid, their applications (for example, whether they were successful),
and the lawyer assigned to them by Legal Aid\textemdash
the latter if aid is granted.
It includes rich information on the applicants\textemdash
such as income, assets, number of dependants and welfare receipts\textemdash
because these factors are used to determine the outcome of the application.

The second dataset is the Re-Offending Database (ROD),
developed and maintained by the
New South Wales Bureau of Crime Statistics and Research (BOCSAR).
From this database, I extract charge-level data on indictable matters.
These include information court outcomes such as fine amounts, guilty pleas,
and incarcerations.
They also include some demographic characteristics of the
defendant, such as gender and age, and details on the court that processed the
case.

BOCSAR was in charge of linking Legal Aid applications to their court
proceedings information across the two datasets.
Most of the cases are linked using a case-specific identifier code.
Whenever this information is lacking, they are linked by matching the date
in which the application for aid was submitted with the court date,
together with the name and date of birth of the applicant/defendant.
For privacy purposes, such identifying information was not included in the
final dataset I received.

In my analysis, I focus on seven case-level outcomes:
``reduced charges'', which indicates if the number
of finalised charges is less than that of initial charges;
``reduced seriousness'', equal to one if a defendant was not
found guilty of their most serious initial charge.
I also look at whether a defendant was incarcerated
(``incarcerated'') and for how long\textemdash the latter both unconditionally
(that is, including zero values) and conditionally on being incarcerated.
Finally, I look at whether the defendant pleaded guilty to their
highest charge, and whether they were fined.

\section{Descriptive Analysis \label{sec:descriptive}}
Table \ref{tab:cov_balance_strict} compares the mean values of key covariates
across treatment groups. For some covariates, I show their unconditional mean
(by treatment group), the share of non-zero values\footnote{
    These variables do not have negative values\textemdash they are bounded at
    zero.
} and the mean of their positive values. I do this for covariates that
have a large share of zero values, which makes their unconditional
means uninformative.
The complete list of covariate names and descriptions are included in the
Appendix \ref{appx:var_des}.

We can infer from Table \ref{tab:cov_balance_strict} that
a mean comparison between the outcomes of
treatment and control groups would likely be affected by selection
bias.
Indeed, applicants whose application for legal aid was denied or terminated
have on average higher incomes, more valuable assets, spend more on housing
and are less likely to have a disability.
These are all factors that increase their probability of facing a rejection
or termination, and these factors are also likely to be affecting
their outcomes\textemdash for instance via how much they are able to
spend on legal fees. They are confounders and would introduce
selection bias into treatment effect estimates unless accounted for.
On the other hand, applicants whose application for legal aid was
denied or terminated have more dependants on average (that is, household members
relying on their income), which decreases their probability of a rejection.
Finally, the two groups are similar across age and share of females.

\begin{table}
\centering
\caption{\label{tab:cov_balance_strict}Covariate balance table for legal aid applicants}
\centering
\fontsize{11}{13}\selectfont
\begin{threeparttable}
\begin{tabular}[t]{lllllll}
\toprule
\multicolumn{1}{c}{ } & \multicolumn{2}{c}{Granted} & \multicolumn{2}{c}{Refused/Terminated} & \multicolumn{2}{c}{Difference} \\
\cmidrule(l{3pt}r{3pt}){2-3} \cmidrule(l{3pt}r{3pt}){4-5} \cmidrule(l{3pt}r{3pt}){6-7}
Variable & Mean & Obs. & Mean & Obs. & Value & p-value\\
\midrule
Gross income (A\textdollar) & 134.40 & 29,207 & 280.73 & 3,946 & -146.33 & 0.0000\\
 & (262.19) &  & (665.18) &  & (10.70) & \\
\hspace{0.5cm}Any gross income (A\textdollar) & 0.34 & 29,207 & 0.44 & 3,946 & -0.09 & 0.0000\\
 & (0.47) &  & (0.50) &  & (0.01) & \\
\hspace{0.5cm}Positive gross income (A\textdollar) & 391.10 & 10,037 & 644.43 & 1,719 & -253.32 & 0.0000\\
 & (315.66) &  & (884.02) &  & (21.55) & \\
Net assets (A\textdollar) & 435.66 & 29,207 & 12,666.64 & 3,946 & -12,230.98 & 0.0000\\
 & (14,873.93) &  & (136,274.12) &  & (2,171.12) & \\
\hspace{0.5cm}Any net assets (A\textdollar) & 0.01 & 29,207 & 0.08 & 3,946 & -0.07 & 0.0000\\
 & (0.11) &  & (0.27) &  & (0.00) & \\
\hspace{0.5cm}Positive net assets (A\textdollar) & 35,345.45 & 360 & 155,225.33 & 322 & -119,879.87 & 0.0000\\
 & (129,463.72) &  & (453,905.55) &  & (26,199.30) & \\
Weekly Housing (A\textdollar) & 44.75 & 29,207 & 77.60 & 3,946 & -32.85 & 0.0000\\
 & (91.96) &  & (126.23) &  & (2.08) & \\
\hspace{0.5cm}Any weekly Housing (A\textdollar) & 0.28 & 29,207 & 0.38 & 3,946 & -0.10 & 0.0000\\
 & (0.45) &  & (0.49) &  & (0.01) & \\
\hspace{0.5cm}Positive weekly Housing (A\textdollar) & 162.03 & 8,067 & 203.74 & 1,503 & -41.71 & 0.0000\\
 & (107.77) &  & (127.02) &  & (3.49) & \\
Motor vehicle (A\textdollar) & 587.54 & 29,207 & 1,921.19 & 3,946 & -1,333.66 & 0.0000\\
 & (3,636.04) &  & (9,287.10) &  & (149.37) & \\
\hspace{0.5cm}Any motor vehicle (A\textdollar) & 0.10 & 29,207 & 0.18 & 3,946 & -0.09 & 0.0000\\
 & (0.30) &  & (0.39) &  & (0.01) & \\
\hspace{0.5cm}Positive motor vehicle (A\textdollar) & 6,095.98 & 2,815 & 10,588.03 & 716 & -4,492.06 & 0.0000\\
 & (10,179.63) &  & (19,595.68) &  & (757.04) & \\
No. of Dependants & 0.15 & 29,207 & 0.23 & 3,946 & -0.08 & 0.0000\\
 & (0.63) &  & (0.73) &  & (0.01) & \\
Female & 0.12 & 29,207 & 0.13 & 3,946 & -0.01 & 0.2246\\
 & (0.33) &  & (0.34) &  & (0.01) & \\
Age & 34.77 & 29,207 & 35.22 & 3,946 & -0.45 & 0.0414\\
 & (12.32) &  & (13.19) &  & (0.22) & \\
Disability & 0.15 & 29,207 & 0.12 & 3,946 & 0.03 & 0.0000\\
 & (0.35) &  & (0.32) &  & (0.01) & \\
\bottomrule
\end{tabular}
\begin{tablenotes}
\item \textit{Notes:} this table reports the mean covariate values by 
  treatment group and their difference, for a selection of covariates. 
  Column 1 reports the mean covariate values for legal aid recipients.
  Column 2 reports the number of observations for this group.
  Column 3 reports the mean covariate values for legal aid applicants to whom 
  aid was either refused, or granted but then terminated.
  Column 4 reports the number of observations for this group.
  Column 5 reports the difference between control and treatment 
  groups\textemdash Columns 1 and 3, respectively. The p-value 
  of this difference is reported in column 6.
  For ``Gross income'', ``Net asstes'', ``Weekly housing'', and 
  ``Motor vehicle'', I report the unconditional mean, the share of 
  positive values and the mean of positive values. 
  As these variables are zero-inflated, their unconditional means are imprecise. 
  Notice also that these variables are bounded at zero. 
  Standard errors are reported between parentheses.
  
\end{tablenotes}
\end{threeparttable}
\end{table}

Inspecting the (unadjusted) mean outcomes for the whole sample
and by treatment group can also be informative.
Table \ref{tab:tab:outcm_bal} reports the mean value for
each outcome. Columns 2 and 3 report the mean values
for the control and treatment groups, respectively, while columns 4 and 5
report the mean differences and their p-value, respectively. Column 6 reports
the mean values for the whole sample, unconditional on treatment status.
It should be moted that the size of the treatment group
(cases where aid was denied or terminated)
is 13.5\% of that of the control group (cases where aid was granted), implying
that the unconditional means underrepresent the treatment group.

\begin{figure}[h]
    \centering
    \subfloat[All]{\label{fig:unc_jail_hist}\includegraphics[width=.49\linewidth]{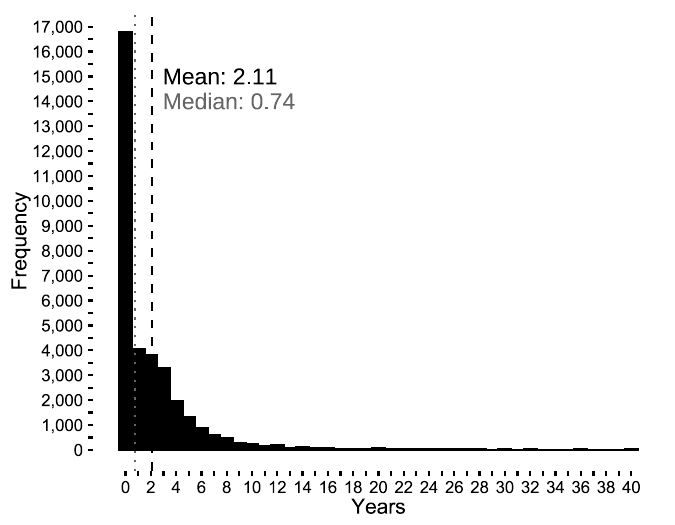}}\hfill
    \subfloat[Strictly Positive]{\label{fig:cond_jail_hist}\includegraphics[width=.49\linewidth]{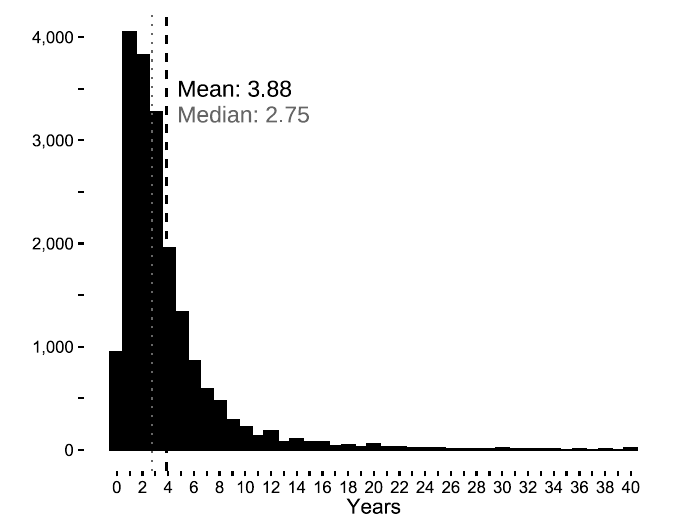}}\par
    \caption{Length of Incarceration Spell}
    \smallskip
    {\footnotesize
        \textit{Notes.} This figure plots incarceration data for the study
        sample, which links legal aid applications (sourced from Legal Aid NSW,
        Crime Division) to their court outcomes (sourced from BOCSAR's ROD).
    }
    \label{fig:jail_hist}
\end{figure}

Inspecting Table \ref{tab:tab:outcm_bal}, we can see that
the share of charges for which defendants are
found guilty does not differ across treatment groups. The same
applies to the probability for a defendant to have their charges reduced.
Most charges are finalised\footnote{
    Finalising a charge means that the defendant is found guilty of the charge.
}
(66\%) and it is relatively rare for the most serious
charge to be withdrawn or dismissed (11\%).
Lawyers manage to reduce the charges of the defendants 57\% of the time
and legal aid lawyers achieve that 2 p.p. more frequently than private lawyers
and self-represented defendants.
The rates of incarceration sit at 57\%, and is 58\% for legal aid cases and
43\% for private and self-represented. Legal aid recipients are expected to
spend 27 months in jail, 6 months longer than their counterparts.
However, conditional on going to jail, legal aid defendants spend 42 months
in custody\textemdash 3 months less than privately- and self-represented
defendants. See Figure \ref{fig:jail_hist} for a visualisation of
all incarceration data (Panel a) and conditional on positive incarceration
spells (Panel b).
This could suggest that private lawyers are better at keeping out of jail
those defendants whose expected time in jail is relatively low.
Indeed, the unconditional incarceration length (which includes zero values)
is higher for legal aid defendants (27 months) versus non-legal aid (21 months).
The share of guilty pleas to the highest charge is high in both
groups\textemdash 65\% and 63\% for legal aid and non-legal aid defendants,
respectively\textemdash and about 3 p.p. higher for legal aid defendants.
Finally, the share of fined defendants is 6\% and, unsurprisingly,
legal aid defendants are less likely to be fined than their
counterparts.

\begin{table}
\centering
\caption{\label{tab:tab:outcm_bal}Balance-type table for outcomes}
\centering
\begin{threeparttable}
\begin{tabular}[t]{llllll}
\toprule
 & $D_i=0$ & $D_i=1$ & Diff. & p-value & All\\
\midrule
Reduced charges & 0.57 & 0.55 & 0.02 & 0.0124 & 0.57\\
 & (0.49) & (0.50) & (0.01) &  & (0.49)\\
Guilty share & 0.66 & 0.66 & 0.00 & 0.8740 & 0.66\\
 & (0.36) & (0.37) & (0.01) &  & (0.36)\\
Reduced seriousness & 0.11 & 0.10 & 0.01 & 0.1907 & 0.11\\
 & (0.31) & (0.30) & (0.01) &  & (0.31)\\
Incarceration (months) & 27.13 & 21.16 & 5.97 & 0.0000 & 26.42\\
 & (44.78) & (44.22) & (0.75) &  & (44.75)\\
Incarceration $>$ 0 (months) & 46.42 & 49.21 & -2.79 & 0.0495 & 46.67\\
 & (50.35) & (56.28) & (1.42) &  & (50.92)\\
Incarcerated & 0.58 & 0.43 & 0.15 & 0.0000 & 0.57\\
 & (0.49) & (0.50) & (0.01) &  & (0.50)\\
Guilty plea to highest charge & 0.65 & 0.63 & 0.03 & 0.0015 & 0.65\\
 & (0.48) & (0.48) & (0.01) &  & (0.48)\\
Fined & 0.05 & 0.11 & -0.06 & 0.0000 & 0.06\\
 & (0.23) & (0.32) & (0.01) &  & (0.24)\\
\bottomrule
\end{tabular}
\begin{tablenotes}
\item \textit{Notes:} this table reports the mean outcome values by 
  treatment group, their difference, and the means for the whole sample. 
  Column 1 reports the mean outcome values for legal aid recipients.
  Column 2 reports the mean covariate values for legal aid applicants to whom 
  aid was either refused, or granted but then terminated.
  Column 3 reports the difference between control and treatment 
  groups\textemdash Columns 1 and 2, respectively. The p-value 
  of this difference is reported in column 4.
  Column 5 reports the means for the whole sample, unconditional on treatment 
  status. Notice that legal aid recipients are overrepresented compared 
  to their counterpart, as the size 
  of the treatment group is 13\% of that of the treatment group, of legal aid 
  recipients.
  Standard errors are reported between parentheses.
  
\end{tablenotes}
\end{threeparttable}
\end{table}

\section{Empirical Strategy \label{sec:methods}}

In this section, I present the empirical strategy used to identify the effect
of legal aid on court outcomes. I use a double machine learning estimator
on administrative data to learn the unknown
treatment assignment function and estimate
the average treatment effect on the treated.
To identify treatment effects, DML requires that the treatment assignment is
\textit{strongly ignorable}\textemdash in the words of
\cite{rosenbaum1983}\textemdash or, in other words, that it is conditionally
\textit{unconfounded} \citep{imbens2004, imbens2015}, that is, that, conditional
on observed covariates, it
(i) does not depend on potential outcomes, (ii) is
probabilistic\textemdash no unit is assigned to treatment or control with
probability one\textemdash and (iii) is individualistic.

Applied to this context, the main threat to assumption (i) is that Legal Aid
might condition its grant decision on predicted case success in ways not
observed by the researcher. This is unlikely for two reasons.
First, for serious criminal matters, Legal Aid does not allow
merit-based rationing\textemdash staff are not permitted to grant or deny aid on
the basis of predicted case success (for example, probability of incarceration).
Second, the factors that are used to make grant decisions are recorded in the
Legal Aid administrative file and used in this study as covariates.
Assumption (ii) is supported empirically
by the estimated propensity score, plotted in Figure \ref{fig:main_ps}.
Finally, assumption (iii)
mainly imposes no crowding externalities, that is, that the Legal Aid budget
has not been binding historically.
Had the Legal Aid budget bound in the past, these rejections would be observable
in the present dataset and labelled as rejected due to a lack of funds.
This is a legitimate reason to deny aid, but the budget allocated to Legal Aid
has never run out thus far and hence it is not recorded in the data.

Section~\ref{sec:irm} presents the Interactive Regression Model (IRM)
estimator of \citet{chern2018}. Like other DML estimators (when treatment is
binary), IRM combines the propensity score and the outcome regressions\textemdash
one per treatment state\textemdash to estimate average treatment effects.
In addition, IRM allows the treatment effects to vary with observed covariates,
meaning it does not impose homogeneous treatment effects.
Intuitively, DML estimators\textemdash IRM included\textemdash
can be viewed as flexible implementations of
the augmented inverse probability weighting (AIPW) estimator
\citep[for example,][]{robins1994,robins1995} (for discussions of how DML relates to
AIPW, see \citeauthor{knaus2022a}, \citeyear{knaus2022a},
\citeauthor{wager2025}, \citeyear{wager2025}, more technical, and
\citeauthor{moccia2024}, \citeyear{moccia2024}, more accessible).
More generally, inverse probability weighting (IPW) targets treatment effects
by modelling the probability of treatment for each unit
(the \textit{propensity score}) and re-weighting observations accordingly.
AIPW augments IPW by also modelling outcomes, resulting in the
double-robustness property\textemdash briefly discussed in Section~\ref{sec:irm}
below.
While this approach allows me to identify both
ATE and ATT, I focus on the ATT in my analysis. Indeed, the ATT can be
estimated more reliably in cases like the present,
where the treatment group is much smaller than the control group.
Intuitively, in such cases propensity score estimators can more easily find
propensity-score matches for treated units in the wider control group.


Before delving into the technical part, stating the key idea concisely
can be helpful. Legal Aid decides on aid allocation based on the information
contained in a form. All this information is included in my dataset\footnote{
    The only field of the application form but not included in the dataset
    is the applicant's free-text statement on the circumstances of the alleged
    offence.
    This not is an input in the decision to provide aid for serious crimes,
    as indictable-crimes cases automatically pass the Merit test (see
    Section \ref{sec:eligibility}).
}.
The allocation of aid follows sharp rules only partly, as they are
complemented by a standards-based approach.
My claim is that the flexibility of double machine learning with random
forests allows me to credibly recover the treatment assignment function and
hence identify both ATE and ATT while addressing the two main threats
to identification. First, such flexibility effectively tackles
potential bias due to functional-form misspecification.
Second, the complete access to
the form data used for assigning treatment
greatly reduces omitted variable bias concerns.

\subsection{Interactive Regression Model \label{sec:irm}}
\cite{chern2018} applies the general DML theory to four different models, two
of which use instrumental variables. Both of the remaining two models can be
used in the present application: the Partially Linear Regression model (PLR)
and the Interactive Regression Model (IRM).
I focus on the IRM as (i) it subsumes PLR as a special case, and (ii) it
explicitly estimates a propensity score, which is useful to help unpacking
and interpreting the results. The only advantage of choosing PLR over IRM
is statistical efficiency and this is not a prime concern in the
present application.

Given a binary treatment variable, the Interactive Regression Model estimates
fully heterogeneous average treatment effects. Here, ``fully heterogeneous''
means that each of the two response curves\textemdash
that is, the potential outcomes as a function of $\mathbf{X}$\textemdash
is allowed to be a different nonparametric function.
Vectors $\mathbf{Y}$ and $\mathbf{D}$ are modelled such that
\begin{eqnarray}
    \mathbf{Y} = g_0(\mathbf{D}, \mathbf{X}) + \mathbf{U},
    & \mathbb{E}(\mathbf{U} \mid \mathbf{X}, \mathbf{D}) = \mathbf{0},
    \label{eq:dml_y}\\
    \mathbf{D} = m_0(\mathbf{X}) + \mathbf{V},
    & \mathbb{E}(\mathbf{V} \mid \mathbf{X}) = \mathbf{0}
    \label{eq:dml_d}.
\end{eqnarray}

This model allows me to estimate not only the ATE, here written as:
$$
    \tau_{ATE}=\mathbb{E}\left[g_0(\mathbf{1}, \mathbf{X})-g_0(\mathbf{0}, \mathbf{X})\right]
$$
but also the ATT:
$$
    \tau_{ATT}=\mathbb{E}\left[g_0(\mathbf{1}, \mathbf{X}) - g_0(\mathbf{0}, \mathbf{X}) \mid \mathbf{D}=\mathbf{1}\right] .
$$

Variables $\mathbf{X}$ are allowed to affect both the outcome $\mathbf{Y}$ and
the treatment $\mathbf{D}$, each via a different function\textemdash
$m_0(\mathbf{X})$ and $g_0(\mathbf{X})$, respectively.
In particular, $m_0(\mathbf{X})$ is the propensity score and as such, its
estimate can be inspected and used to test assumptions such as the common
support assumption.
Doing so can ease the black-box concerns typical of ML applications.
Moreover, both $m_0(\mathbf{X})$ and $g_0(\mathbf{X})$ are unknown
and likely to be highly non-linear in this application, where the decision
to assign a unit to treatment is made via a mix of sharp rules and standards.
Hence, the benefits of estimating them
flexibly using ML methods \citep[here random forests,][]{athey2019}
are potentially large.

Another useful property of \cite{chern2018}'s DML estimators
is \textit{double robustness}, both weak and strong \citep{wager2025}.
Weak double robustness means that the estimator is \emph{consistent} for the
estimand (here, $\tau_{ATE}$ and $\tau_{ATT}$) if either the outcome
model (\ref{eq:dml_y}) or the treatment model
(\ref{eq:dml_d}) is consistently estimated\textemdash
even if the other is not.
Strong double robustness means that the estimator is constructed so that
first-order sensitivity to errors in the nuisance parameters
($g_0(\cdot)$ and $m_0(\cdot)$) is attenuated:
the leading bias term depends primarily on how large the estimation errors
in the two nuisance components are
\emph{jointly} (in the sense that it scales with their interaction), rather than on
the sum of their separate errors. This is the key reason DML can support
standard large-sample inference for the treatment effect (that is, $n^{-1/2}$-rate
convergence and asymptotic normality) even when $g_0$ and $m_0$ are learned using
flexible methods that may converge more slowly than $n^{-1/2}$, \emph{provided},
roughly, that their estimation errors is sufficiently small in combination
\citep{wager2025, naimi2020}.
In the present application,
particular attention is paid to the treatment model, as the institutional
setting suggests that the variables used to assign funding are observed,
and I include them in the determinants of treatment assignment, $X$.

In my preferred specification, I add court-level fixed effects to Equations
\ref{eq:dml_y} and \ref{eq:dml_d} using a within transformation.
This is to model how cases are assigned
to Legal Aid offices and thus control for potential variation in the
performance of Legal Aid offices\textemdash given as-good-as-random
case allocation within office.
The inclusion of fixed effects is based on ex-ante grounds,
since ex-post they do not seem to be important drivers of the estimates
(see Section \ref{sec:results}).

Finally, it is worth noting that the ML algorithms (including random forests),
are designed for prediction and typically produce biased
estimates of treatment effects and invalid inference.
The insight of DML is not only to use ML algorithms once on the
outcome model and once on the treatment model, but to
combine them with orthogonal scores and cross-fitting
(an efficient form of data-splitting) to remove the
regularisation bias and overfitting that usually make ML estimators unhelpful
in estimation settings.

To implement DML with random forests, I use five cross-fitting folds and
repeat the cross-fitting procedure ten times before aggregating the
results, as suggested in \cite{chern2018}.
To learn $g_0(\mathbf{X})$ and $m_0(\mathbf{X})$, I use
a random forest regression and a random forest classifier, respectively.
In both random forest algorithms, the number of trees is set to 500, the
minimum node size to two and the maximum depth to five.

\subsection{A Note on Intensive Margin Estimates \label{sec:disclaimer}}
It should be noted that treatment effects estimated after conditioning on
post-treatment variables should always be interpreted with caution, as
doing so exposes these estimates to collider bias
\citep[for example, see][]{collider}.
This is the case of the treatment effects on the length of
incarceration spells when the spells are positive\textemdash
the intensive margin.

Here, this means that factors that are not modelled and
are correlated with both incarceration length and the treatment variable
could introduce bias.
For the intensive margin estimate to be unbiased, there would need to be
no remaining unmodelled factor that simultaneously drives both
(i) the probability of being denied aid (the focus of Section \ref{sec:pref})
or whether the lawyer is public or private
(the focus of Section \ref{sec:priv_pub}), and
(ii) the length of the resulting sentence.
I further discuss this issue and the argument for including these estimates in
Section \ref{sec:discuss}.

\section{Results \label{sec:results}}
In this section, I start by presenting my preferred estimates\footnote{
    All DML estimates are calculated via R package \cite{DoubleML}
} for the
effect of denying legal aid on court outcomes, among legal
aid applicants charged with serious crimes.
I then ``unpack'' the results by (i) plotting the estimated propensity score by
treatment group; (ii) showing a propensity-adjusted covariate-balance table;
(iii) plotting variable importance, which shows which covariates best predict
being denied aid; (iv) using partial dependence plots (PDP) to
further explore the role played by income and assets in being denied aid; and
(v) conducting a formal omitted-variable bias analysis.
Finally, I discuss the generaliseability of my estimates to the whole population
of legal aid applicants charged with serious crimes. This includes
applicants whose aid application was initially granted but then terminated, and
successful applicants that are represented by publicly-funded private lawyers,
that is, \textit{panel lawyers}.

\subsection{Preferred Estimates \label{sec:pref}}
The preferred estimates are reported in Table
\ref{tab:tab:irm_one_ATT_inhouse_noterm}.
They involve sample restriction and modelling choices addressing two key
sources of concern.
First, terminated cases could introduce bias. For instance,
public lawyers could be dropping misbehaving clients, who would then
have to hire a private lawyer or self-represent, hence biasing the outcomes
of the treated group downward.
Second, while case assignment is as good as random among in-house
lawyers within the same office,
panel lawyers can self-select into particular cases based on their
expertise, hence introducing unmodelled heterogeneity.
I address these issues by running the DML analysis after dropping cases
(i) where the application for aid was initially granted, but then terminated,
and (ii) where defendants who were granted aid were represented by panel
lawyers. Additionally, I introduce court-level
fixed effects via a within transformation to model the random assignment
of cases to legal aid lawyers at the office level.

\begin{table}
  \centering
  \caption{\label{tab:tab:irm_one_ATT_inhouse_noterm}DML: ATT of Denying Aid on Court Outcomes
    in a sample without terminated grants nor panel lawyers}
  \centering
  \begin{threeparttable}
    \begin{tabular}[t]{lrrrrrr}
      \toprule
                                       & Coef.  & s.e.  & l-CI   & r-CI   & p-value & Obs.  \\
      \midrule
      Reduced charges                  & -0.005 & 0.013 & -0.030 & 0.021  & 0.7250  & 16854 \\
      Reduced seriousness              & -0.004 & 0.005 & -0.015 & 0.006  & 0.3992  & 14875 \\
      Incarcerated - extensive         & -0.081 & 0.012 & -0.106 & -0.057 & 0.0000  & 16852 \\
      Incarceration (mth.)             & -0.124 & 0.830 & -1.750 & 1.502  & 0.8816  & 16854 \\
      Incarceration - intensive (mth.) & 5.586  & 1.549 & 2.550  & 8.621  & 0.0003  & 8753  \\
      Guilty plea to highest charge    & -0.046 & 0.010 & -0.065 & -0.027 & 0.0000  & 16854 \\
      Fined                            & 0.048  & 0.008 & 0.033  & 0.063  & 0.0000  & 16846 \\
      \bottomrule
    \end{tabular}
    \begin{tablenotes}
      \item \textit{Notes:} The table presents Double Machine Learning ATE
      estimates for the effect of denying aid on case-level court outcomes
      among applicants charged with serious crimes.
      The IRM model is applied to all cases where aid was not terminated and
      where the defendant was not represented by a panel lawyer.
      The chosen ML method is Random Forests.
      Finally, court-level fixed effects are included via a within transformation.
    \end{tablenotes}
  \end{threeparttable}
\end{table}



\begin{table}
  \centering
  \caption{\label{tab:tab:irm_one_inhouse_noterm_norep} DML: ATE of Denying Aid on Court Outcomes
    in a sample without terminated grants, panel lawyers, or self-represented defendants}
  \centering
  \begin{threeparttable}
    \begin{tabular}[t]{lrrrrrr}
      \toprule
                                       & Coeff. & s.e.  & l-CI   & r-CI   & p-value & Obs.  \\
      \midrule                                                                              \\ [-5.31ex] \\  \multicolumn{1}{l}{\textit{Panel A: ATT}} \vspace{2pt}\\
      Reduced charges                  & -0.015 & 0.013 & -0.041 & 0.011  & 0.2546  & 16339 \\
      Reduced seriousness              & -0.009 & 0.005 & -0.019 & 0.002  & 0.1016  & 14416 \\
      Incarcerated - extensive         & -0.097 & 0.013 & -0.123 & -0.071 & 0.0000  & 16338 \\
      Incarceration (mth.)             & -0.474 & 0.850 & -2.141 & 1.193  & 0.5772  & 16339 \\
      Incarceration - intensive (mth.) & 5.791  & 1.614 & 2.628  & 8.954  & 0.0003  & 8734  \\
      Guilty plea to highest charge    & -0.050 & 0.010 & -0.069 & -0.031 & 0.0000  & 16339 \\
      Fined                            & 0.066  & 0.010 & 0.047  & 0.084  & 0.0000  & 16338 \\
      \midrule                                                                              \\ [-5.31ex] \\  \multicolumn{1}{l}{\textit{Panel B: ATE}} \vspace{2pt}\\
      Reduced charges                  & -0.016 & 0.011 & -0.039 & 0.006  & 0.1562  & 16339 \\
      Reduced seriousness              & -0.008 & 0.004 & -0.016 & 0.000  & 0.0519  & 14416 \\
      Incarcerated - extensive         & -0.125 & 0.010 & -0.146 & -0.105 & 0.0000  & 16338 \\
      Incarceration (mth.)             & -1.799 & 0.705 & -3.181 & -0.418 & 0.0107  & 16339 \\
      Incarceration - intensive (mth.) & 5.485  & 1.464 & 2.616  & 8.355  & 0.0002  & 8734  \\
      Guilty plea to highest charge    & -0.066 & 0.007 & -0.079 & -0.053 & 0.0000  & 16339 \\
      Fined                            & 0.071  & 0.009 & 0.053  & 0.089  & 0.0000  & 16338 \\
      \bottomrule
    \end{tabular}
    \begin{tablenotes}
      \item \textit{Notes:} The table presents Double Machine Learning ATT and
      ATE estimates for the effect of denying aid on case-level court outcomes
      among applicants charged with serious crimes.
      The IRM model is applied to all cases where aid was not terminated and
      where the defendant was not represented by a panel lawyer
      nor self-represented.
      The chosen ML method is Random Forests.
      Finally, court-level fixed effects are included via a within transformation.
    \end{tablenotes}
  \end{threeparttable}
\end{table}

The estimated effects on ``reduced charges'' (-0.5 p.p.),
``reduced seriousness'' (-0.4 p.p.), and ``incarceration'' (-0.12 months) are
small and statistically insignificant\textemdash they are precise null
effects.
The other coefficients are statistically significant and precise.
Denying legal aid to a defendant reduces their chances of being incarcerated
by 8.1 p.p.
It decreases their probability of pleading guilty to the highest charge by
4.6 p.p. and increases their chances of being fined by 4.8 p.p.
Finally, denying legal aid to those who are incarcerated is linked to a
5.6 months increase in the length of their incarceration spell.

\subsection{Private vs. Public \label{sec:priv_pub}}
Once I drop self-represented defendants, I can focus on the
effect of hiring a private lawyer, where being represented by a public lawyer
is the baseline. Here, it is useful to note that the kind of
private lawyer hired by someone who had a positive chance to qualify for
aid would not be an expensive one (see \textit{Weekly gross income} in Table
\ref{tab:cov_balance_strict}), but rather one cheaper than average.

The treatment effect estimates for the effect of hiring a private lawyer
(Table \ref{tab:tab:irm_one_inhouse_noterm_norep})
are consistent with and slightly larger than
those for the effect of denying aid (Section \ref{sec:pref}).
In particular, privately represented defendants are 9.7 p.p. less likely to be
incarcerated and 5 p.p. less likely to plead guilty of their most serious
charge.
Comparing ATTs and ATEs in Table \ref{tab:tab:irm_one_inhouse_noterm_norep},
panels A and B, respectively,
it should be noted that the average treatment effects are larger than
the average treatment effects on
the treated, implying that the untreated population\textemdash that is, those who
were granted aid and are hence poorer\textemdash
would reap even larger benefits from private lawyers
if they could (counterfactually) afford them.

\subsection{Estimated Propensity Score}
\begin{figure}[!hp]
    \includegraphics[width=\textwidth]{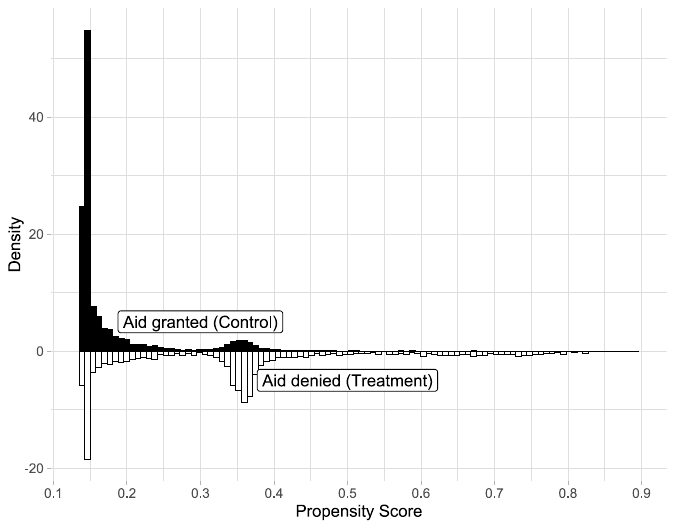}
    \caption{Propensity score for the preferred estimates}
    \label{fig:main_ps}
\end{figure}

To better investigate the preferred treatment effect
estimates, I start by plotting their estimated propensity score.
As Figure \ref{fig:main_ps} shows, first, there are no observations with a
propensity score equal to or close to zero or one.
This provides evidence in favour of the \textit{common support} assumption,
which is required to interpret the results as causal.

Second, the densities of the propensity score across
the treatment and control groups overlap when the propensity is
between zero and 0.45, circa.
After 0.45, there is low overlap, as there exist some treated applicants
(those who are denied aid) that are unlikely to receive aid. While this is
to be expected, if these observations were an important driver of our results,
it would be a source of concern.
Hence, I re-estimate the treatment effects after dropping cases with a
propensity score greater than 0.45.
The results are consistent with the preferred ones, providing evidence that
the latter are not driven by cases with low overlap
(see Table \ref{tab:tab:irm_one_ATT_ps045}).
It should be noted that this test is somewhat informal, as it
produces (statistically) consistent estimates of the treatment effect
but does not produce valid standard error due to selection issues.

\begin{table}
  \centering
  \caption{\label{tab:tab:irm_one_ATT_ps045}DML: ATT of Denying Aid on Court Outcomes
    when the propensity score is 0.45 or less}
  \centering
  \begin{threeparttable}
    \begin{tabular}[t]{lrrrrrr}
      \toprule
                                       & Coef.  & s.e.  & l-CI   & r-CI   & p-value & Obs.  \\
      \midrule
      Reduced charges                  & -0.014 & 0.010 & -0.034 & 0.006  & 0.1649  & 15861 \\
      Reduced seriousness              & -0.004 & 0.007 & -0.017 & 0.009  & 0.5626  & 14106 \\
      Incarcerated - extensive         & -0.089 & 0.010 & -0.108 & -0.070 & 0.0000  & 15859 \\
      Incarceration (mth.)             & -0.362 & 0.779 & -1.888 & 1.165  & 0.6423  & 15861 \\
      Incarceration - intensive (mth.) & 5.563  & 1.464 & 2.694  & 8.432  & 0.0001  & 8487  \\
      Guilty plea to highest charge    & -0.053 & 0.010 & -0.072 & -0.033 & 0.0000  & 15861 \\
      Fined                            & 0.053  & 0.007 & 0.039  & 0.067  & 0.0000  & 15853 \\
      \bottomrule
    \end{tabular}
    \begin{tablenotes}
      \item \textit{Notes:} The table presents Double Machine Learning ATT
      estimates for the effect of being denied aid on case-level court outcomes.
      The IRM model is applied to
      cases where the estimated propensity score is lower or equal to 0.45. This is
      the interval of the propensity score with stronger overlap.
      The chosen ML method is Random Forests.
    \end{tablenotes}
  \end{threeparttable}
\end{table}

I acknowledge that changing the restrictions on the sample for sensitivity
testing inevitably changes the target population
and causal parameter. However, I find it to be a fruitful approach for testing
the sensitivity of the preferred estimates to the removal of observations
that are suspected to be driving the results\textemdash in circumstances
where one would not want this to be the case.

\subsection{Further Sensitivity Analysis}
To further explore the underlying mechanics of the IRM model,
I produce a balance table showing the difference in covariate means across
treatment groups before and after propensity-score weighing (see
Table \ref{tab:ps_bal_tab_main} in the Appendix \ref{appx:other_sens}).
This is only one of the two forms of balancing used by DML so it
\textit{partially} reflects the ability of the model to adjust for differences
across treatment groups.

Table \ref{tab:ps_bal_tab_main} shows an improvement in balance after
adjusting via DML, particularly along the income dimension.
Asset-related variables\textemdash summarised by the \textit{Assessable Net
    Assets} covariate\textemdash are harder to balance, and so is the
private-submission indicator.
On the net assets dimension, this is because only a small fraction of values
are positive (3\% circa), and these include some extreme values.
While zero values are quite evenly spread across the treatment groups
(Table \ref{tab:tab:pos_ASS_NET_ASSETS_main}),
the few extreme-valued assets are among the denied applications (treated group)
and generate a wide gap across treatment groups
(see Table \ref{tab:tab:assets_stats} in the Appendix).
On the private-submission dimensions, this is because
most of those who were granted aid did not submit their application for aid via
a private lawyer (99\%), while those who were denied aid were almost as likely
to have sent it via a private lawyer (56\%) than via a public one (see
Table \ref{tab:tab:prv_sub_main} in the Appendix).

With the above in mind,
I address concerns about any remaining imbalance of the
net-assets and net-income variables biasing the results.
I do this by re-running the preferred model while
dropping applications where the applicant earns strictly more than
A\$ 600 per week (\textit{assessable net income}) and has assessable net assets
valued at strictly more than A\$ 1,000.
In particular, these restrictions are imposed to check whether people
with outlier levels of income and assets are driving the treatment effect
estimates.
I find that
imposing these restrictions does not affect the preferred coefficient estimates
qualitatively, nor affects the patterns of statistical significance
(see Table \ref{tab:tab:irm_one_ATT_600_main} in the Appendix).

\begin{figure}[h]
    \centering
    \includegraphics[width=\linewidth]{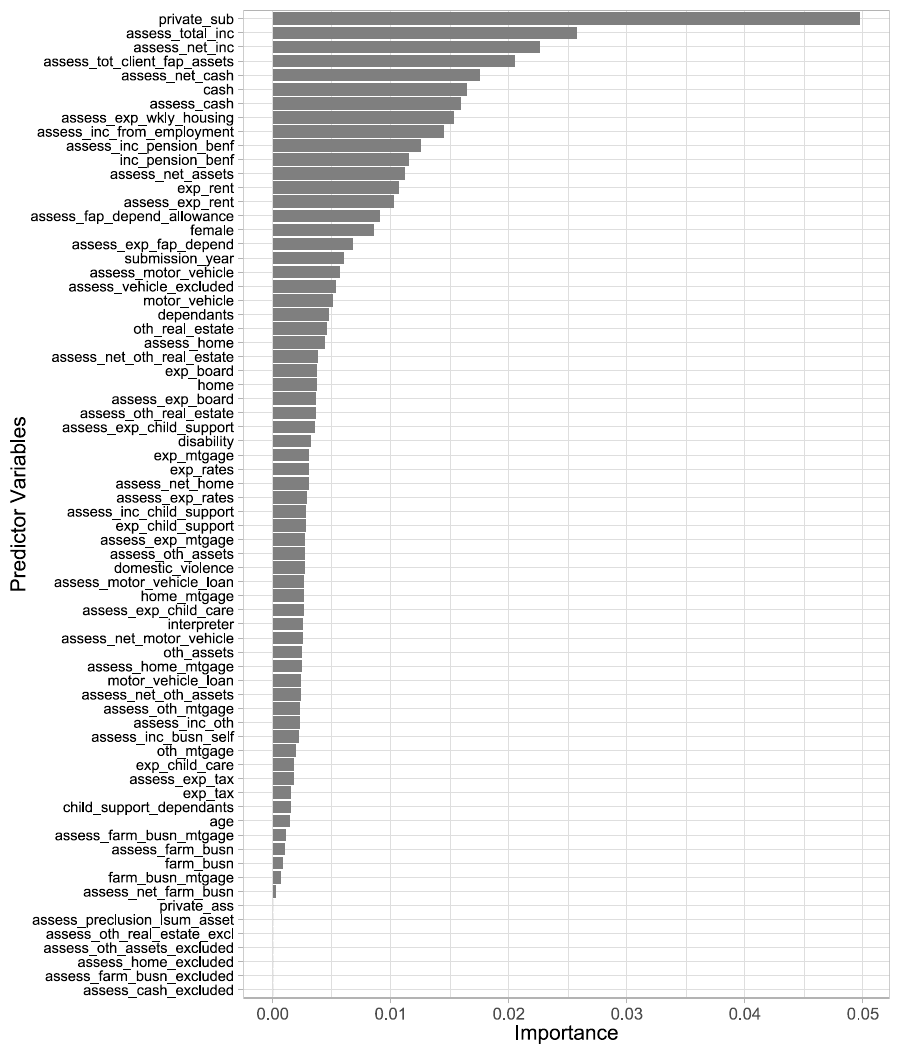}
    \caption{
        \label{fig:var_impt}
        Variable importance of predictors of legal aid denial
    }
    \begin{minipage}{\textwidth}
        \footnotesize
        \vspace{2pt}
        \textit{Notes:}
        This variable importance plot visualises the relative importance of
        the preferred model's covariates in predicting the likelihood
        of a legal aid denial.
        The importance scores reflect each covariates's contribution to the
        model's predictions, with higher values indicating stronger influence.
        It was computed using a permutation-based approach via package
        \cite{ranger}.
    \end{minipage}
\end{figure}

\begin{figure}[h]
    \centering
    \includegraphics[width=\textwidth]{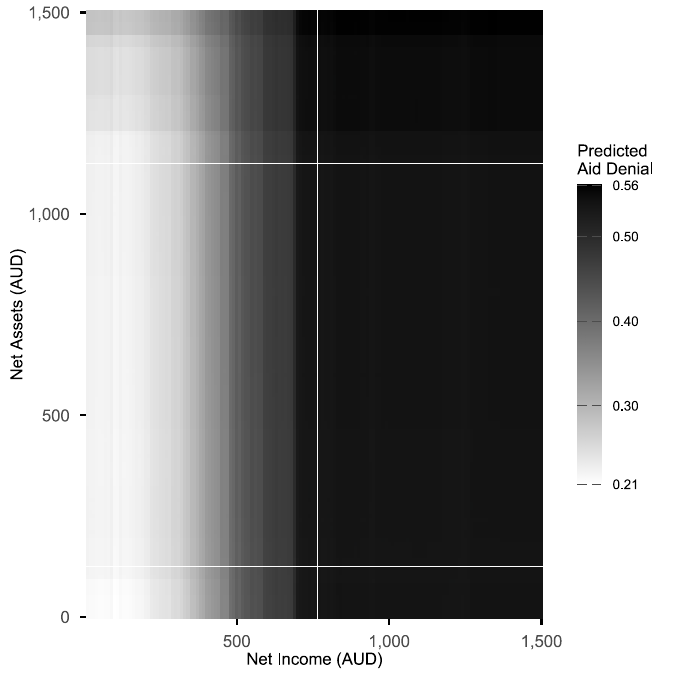}
    \caption{Predicted probability of being denied aid}
    \label{fig:biv_pdp_square}
    \begin{minipage}{\textwidth}
        \footnotesize
        \emph{Notes:} This bivariate partial dependence plot (PDP)
        illustrates how the model's predicted probability of having a legal aid
        application denied changes as both \emph{Assessable Net Income}
        (horizontal axis) and \emph{Assessable Net Assets} (vertical axis)
        vary. The greyscale shows higher denial probabilities in darker shades
        and lower ones in lighter shades.
        This plot is based on estimates from the preferred model
        run on the main data before the within transformation,
        for ease of interpretation.
    \end{minipage}
\end{figure}

I also note that assets play a smaller role than income in predicting
who is denied aid. This is reflected by: (i) the variable importance
plot (Figure \ref{fig:var_impt}), a standard Random Forests metric that here
captures the relative importance of the model's covariates
in predicting the likelihood of a legal aid denial;
and (ii) by a bivariate partial dependence plot of
\textit{Assessable Net Income} and
\textit{Assessable Net Assets}, Figure \ref{fig:biv_pdp_square}, which
illustrates how the model's predicted response varies jointly with
changes in these two covariates, while keeping other covariates
fixed\footnote{This plot is based on estimates from the preferred model
    run on the main data before the within transformation. This is done
    for ease of interpretation.
    The version that is based on the preferred model run on
    the main data after the within transformation has been omitted
    as centring the variables makes interpreting the plot challenging,
    particularly given the presence of extreme outliers.} (for a greater
focus on the relationship between income and aid denial, see Figure
\ref{fig:pdp_inc}).
Specifically, asset variables only appear once in the top ten predictors of
aid denial, at the fourth position, and such variable quantifies the assets
of the financially-associated person (FAP) of the applicant\textemdash their
significant other.
Instead, total income and net income both rank higher and, more broadly,
income-related variables are more represented among the top ten predictors
(four).
Moreover, the bivariate partial dependence plot shows that the
probability of aid denial increases drastically as net income
increases\textemdash the plot's squares darken steadily as we move to the
right\textemdash but barely changes as net assets
increase\textemdash the squares keep the same shade as we move vertically.

Another concern relates to the anomaly that the COVID-19 pandemic represented,
and how it affected the functioning of the criminal justice system.
Since my sample includes the COVID-19 pandemic years, I report
in Appendix \ref{appx:other_sens},
Table \ref{tab:tab:irm_one_ATT_inhouse_noterm_nocovid},
results estimated on a sample dropping grants that have a submission year
above 2019. They are consistent with the main findings.

Furthermore, one may wonder whether the estimated effects remain stable after
focusing on those applications with a higher propensity score, that is, pertaining
defendants that failed the income test, the asset test, or both.
I show in Appendix \ref{appx:other_sens},
Table \ref{tab:tab:irm_one_ATT_fail_inc_assets}, that this is the case\textemdash
the findings are largely consistent although less precise due to the
smaller sample.

Lastly, as a sanity-check, I re-estimate the treatment effects using a
conventional parametric AIPW estimator\footnote{
    I do this in R, via the \cite{aipw} package
}, and include only what I deem to
be the most theoretically relevant covariates (I don't choose them based on
variable importance to avoid selection issues). As standard for parametric
AIPW with binary treatment,
I choose a logistic regression for the propensity score and
for the outcome regression, a logistic regression when the outcome is binary and
an OLS regression when not. The resulting estimates are similar in magnitude
and statistical significance to our main DML results, alleviating concerns related
to issues in the implementation of DML.
This set of results is reported in Table \ref{tab:aipw},
Appendix \ref{appx:other_sens}, together with a complete list of the covariates
included.

\subsection{Sensitivity to Unobserved Confounding\label{sec:sens_longstoryshort}}

In this paper, I claim that all factors that are considered by Legal Aid
officers when making aid decisions are included as covariates in the
analysis.
It is logically possible though, that a factor may not be captured well
by this study dataset, and that the included covariates are not good
proxies for it. \textit{If} that factor were also a confounder\textemdash
affecting both the treatment and the outcome variables\textemdash
then its absence from the covariate list
would bias the estimated treatment effects.
Thanks to \cite{chernozhukov2022longstoryshort},
we can explore this possibility formally\textemdash and we do so below\footnote{
    Given R's version of the package is still in early development,
    we rely on Python's version of the package \citep[see][]{DoubleML2022}.
}.
This sensitivity analysis tool parameterises the strength of an unobserved
confounder (or set of confounders) using two measures akin to
partial-$R^2$ for OLS, but adapted to the DML context:
(i) $cf_y$, capturing the fraction of residual variation in the outcome that
could be explained by unobserved confounding, and
(ii) $cf_d$, capturing the fraction of residual variation in the treatment that
could be explained by unobserved confounding.

A further important parameter is the ``degree of adversity'', $\rho\in[-1,1]$,
which measures how strongly omitted confounding is correlated across the
outcome and treatment channels. More formally, $\rho$ is the correlation
between the omitted-variable errors of the
outcome regression and of the treatment weighting term (the Riesz
representer).
Holding the strength of confounding fixed, $(cf_y,cf_d)$,
the magnitude of the resulting omitted-variable bias increases with $\rho^2$.
Specifically,
when the two channels are weakly correlated, $\rho^2$ is close to zero,
minimising the impact of the omitted confounding on the bias of the
treatment-effect estimate for a given confounding strength\textemdash
\cite{chernozhukov2022longstoryshort} calls this ``amicable confounding''.
When, the channels are strongly correlated, $\rho^2$ is close to one,
maximising the impact of the omitted confounding on the bias of the
treatment-effect estimate for a given confounding strength\textemdash
also called ``adversarial'' confounding.
Please refer to \cite{chernozhukov2022longstoryshort} for a more technical
explanation.

In the present application, I produce several unobserved-confounding scenarios
by using observed covariates identified as potential strong confounders:
(i) the dummy variable
indicating whether the
aid application was submitted via a private lawyer, $private\_sub$, and
(ii) the set of all income variables.
The rationale for using (i) as a benchmark
is that submission via a private lawyer may affect the
selection channel\textemdash either by signalling access to private funding
(raising the probability of denial), or by improving the
application (lowering the probability of denial).
Indeed, we know that $private\_sub$ strongly predicts aid denial thanks to
the variable importance plot (Figure \ref{fig:var_impt}).
Moreover,
even before any legal aid decision is made, having a private lawyer involved
early can affect factors such as bail applications, charge negotiations, and
plea timing, all of which can change the probability of jail\textemdash
the outcome channel.
Finally, submitting via a private lawyer may proxy unobserved traits like
conscientiousness or the ability to engage with the legal process,
which may affect both the treatment assignment and court outcomes.

The rationale for (ii) is that income mechanically affects the chances of
receiving aid, and it may also affect incarceration risk
through correlated factors like the ability to hire private
representation.
This use of observed covariates as benchmarks for unobserved confounding is
helpful to understand what bias scenarios are of practical relevance.
Hence, I use \cite{DoubleML2022} to
estimate confounding strength $(cf_d,cf_y)$ and the
degree of adversity $\rho$ for benchmark variable sets (i) and (ii).
In other words, I
calculate and plot how omitting from the set of covariates a confounder as
strong as $private\_sub$ or all the income variables
would impact my key treatment-effect estimate, that is, the effect of
aid denial on the probability of incarceration.

\begin{figure}[htb]
    \centering
    \thispagestyle{empty}
    \hspace*{-6mm}
    \begin{tikzpicture}[every node/.style={inner sep=0, outer sep=0}]

        \node (privsub_theta_rhohat) {\includegraphics[width=0.5\linewidth]{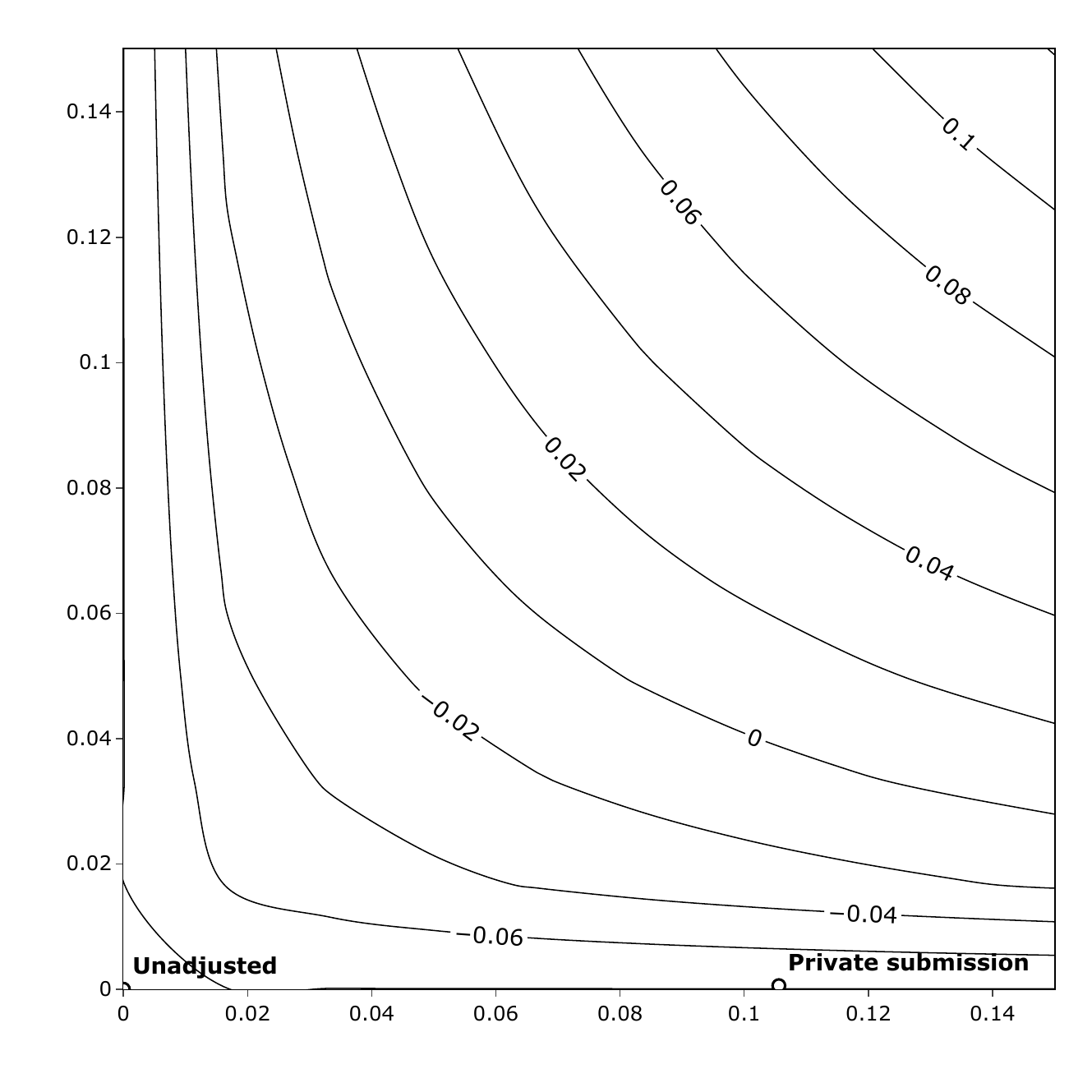}};
        \node[right=0pt of privsub_theta_rhohat] (privsub_ci_rhohat)
        {\includegraphics[width=0.5\linewidth]{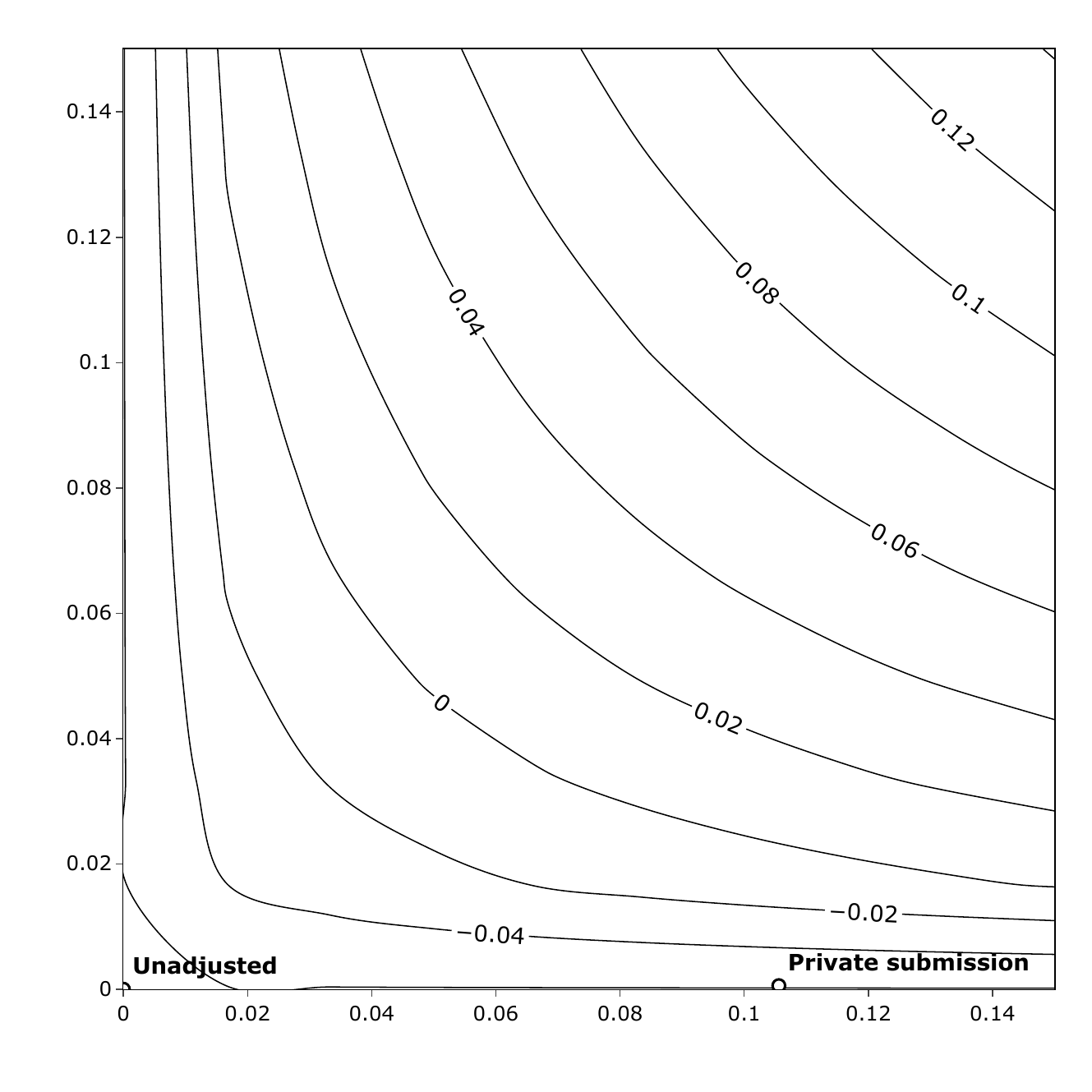}};

        \node[below=0pt of privsub_theta_rhohat] (privsub_theta_rho1)
        {\includegraphics[width=0.5\linewidth]{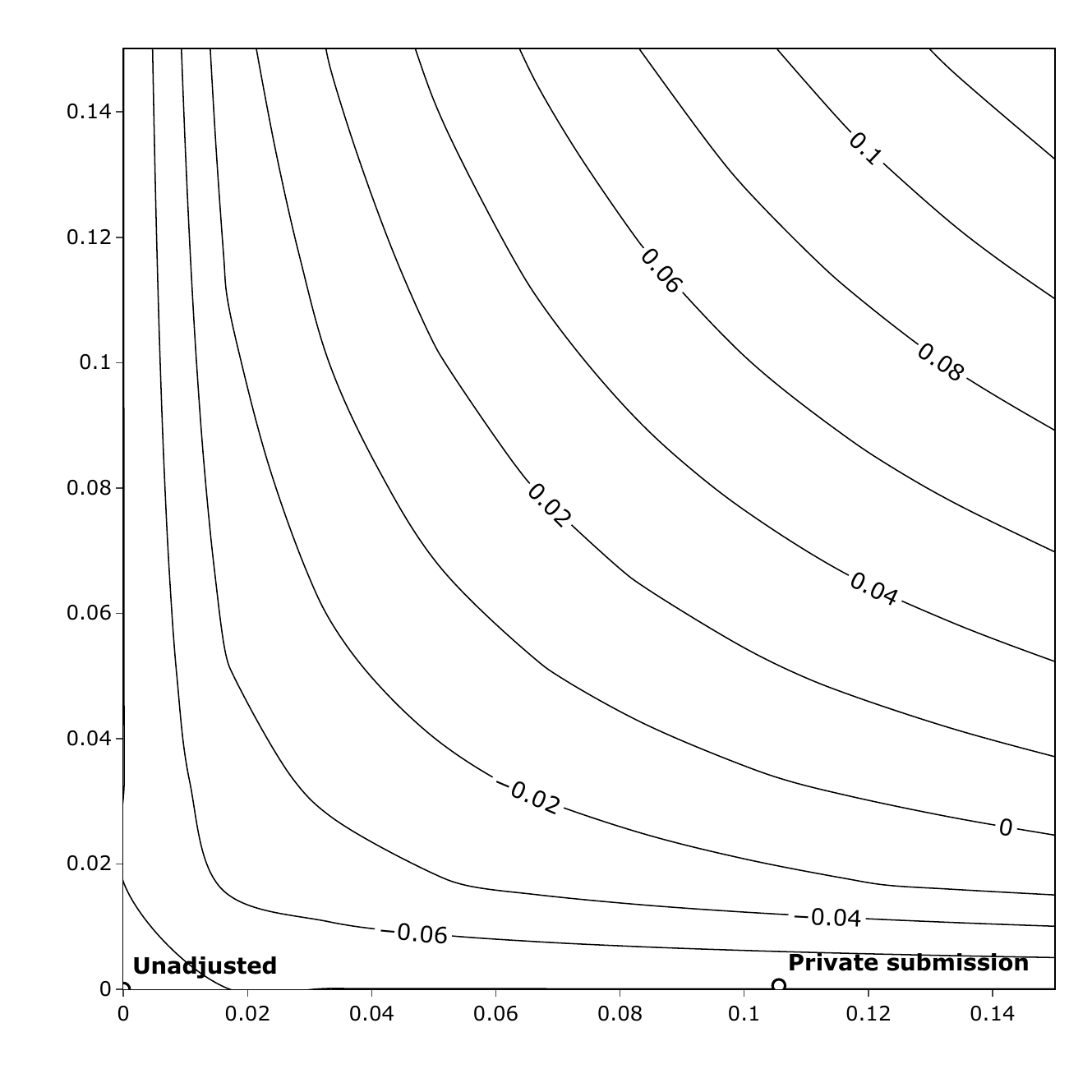}};
        \node[right=0pt of privsub_theta_rho1] (privsub_ci_rho1)
        {\includegraphics[width=0.5\linewidth]{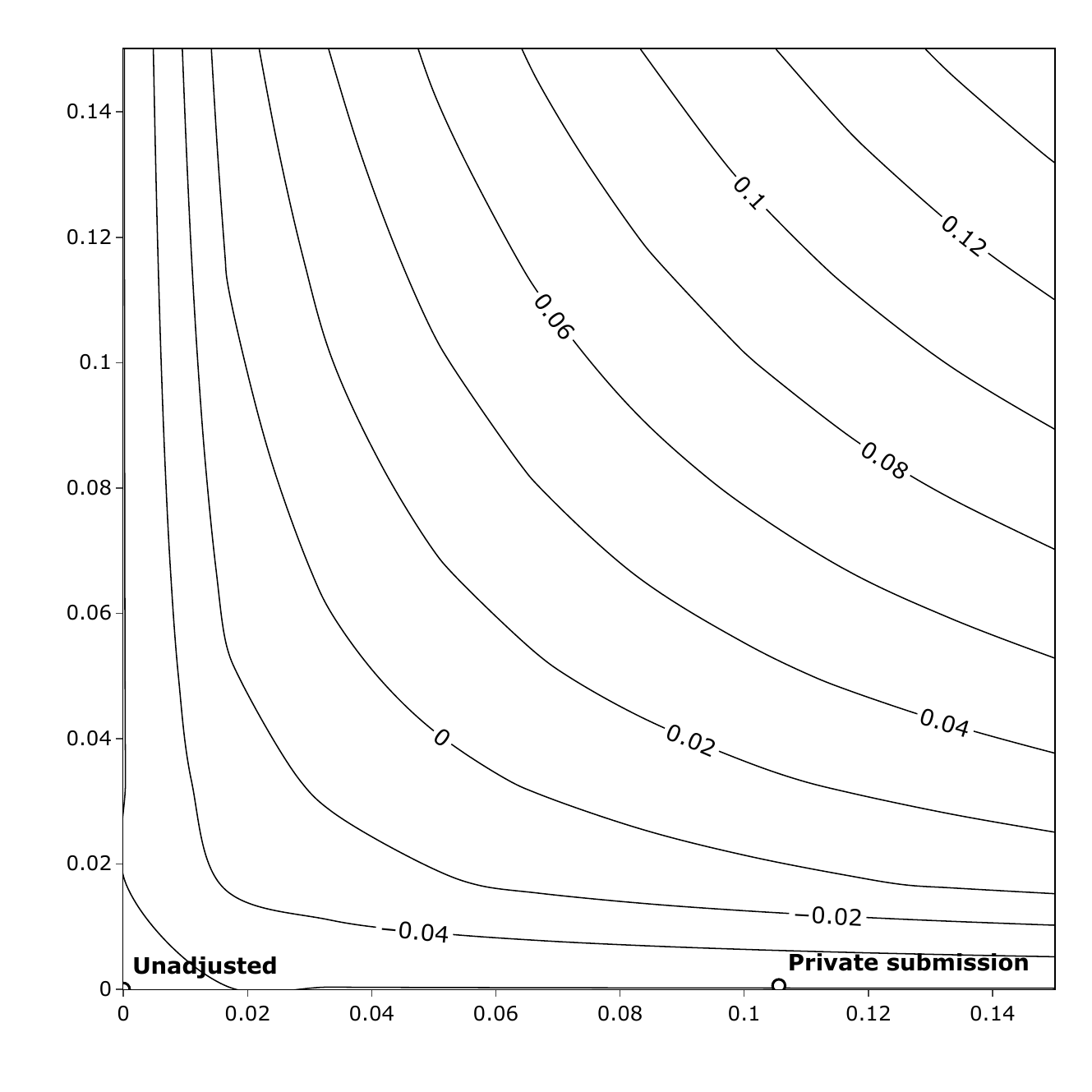}};

        \node[above=2mm of privsub_theta_rhohat, font=\bfseries] {Bound on treatment effect};
        \node[above=2mm of privsub_ci_rhohat, font=\bfseries] {Bound on CI};

        \node[left=8mm of privsub_theta_rhohat, rotate=90, anchor=center, font=\bfseries, align=center]
        {As adversarial as \textit{private submission},\\$\bm{\rho = -0.934$}};
        \node[left=8mm of privsub_theta_rho1, rotate=90, anchor=center, font=\bfseries, align=center]
        {Most adversarial case,\\$\bm{\rho = 1$}};

        \node[below=0mm of privsub_theta_rho1, xshift=0.25\linewidth, font=\small]
        {Confounding strength w.r.t.\ treatment};
        \node[left=0mm of privsub_theta_rho1, yshift=0.5\linewidth, rotate=90, font=\small]
        {Confounding strength w.r.t.\ outcome};

    \end{tikzpicture}
    \caption{Sensitivity contours using $private\_sub$ as the benchmarking confounder.}
    \label{fig:sens_privsub_2x2}
    \begin{minipage}{\textwidth}
        \footnotesize
        \emph{Notes:}
        The left column reports bounds on
        $\theta$ and the right column reports bounds on the upper limit of the 95\% CI.
        Axes give $(cf_d,cf_y)$, the treatment- and outcome-side confounding strengths.
        Rows vary the ``degree of adversity'', $\rho$.
        ``Unadjusted'' is $(cf_d,cf_y)=(0,0)$ and ``Private submission'' is the benchmark point implied by omitting \textit{private\_sub}.
        See Section \ref{sec:sens_longstoryshort}.
    \end{minipage}
\end{figure}

\begin{figure}[htb]
    \centering
    \thispagestyle{empty}
    \hspace*{-6mm}
    \begin{tikzpicture}[every node/.style={inner sep=0, outer sep=0}]

        \node (inc_theta_rhohat) {\includegraphics[width=0.5\linewidth]{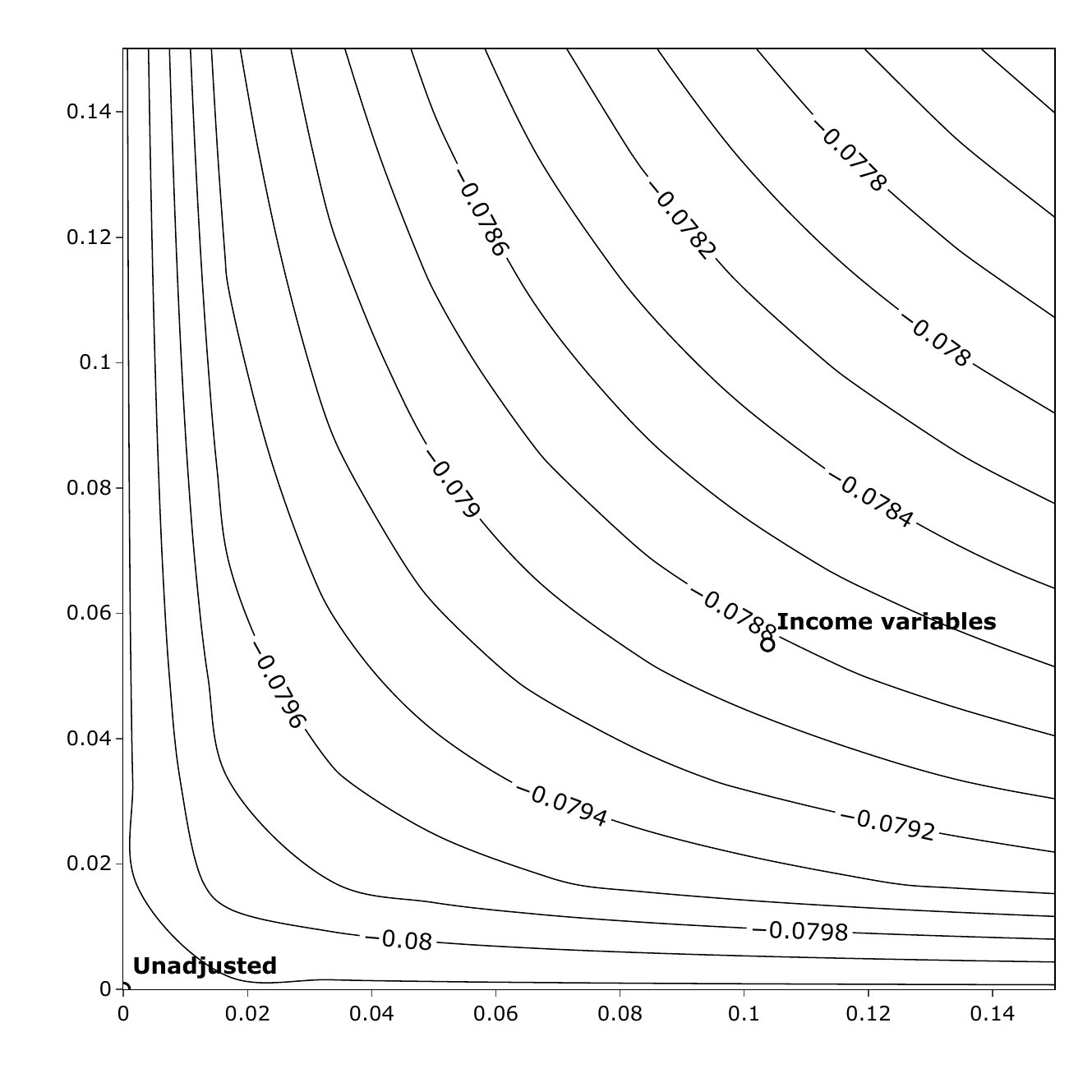}};
        \node[right=0pt of inc_theta_rhohat] (inc_ci_rhohat)
        {\includegraphics[width=0.5\linewidth]{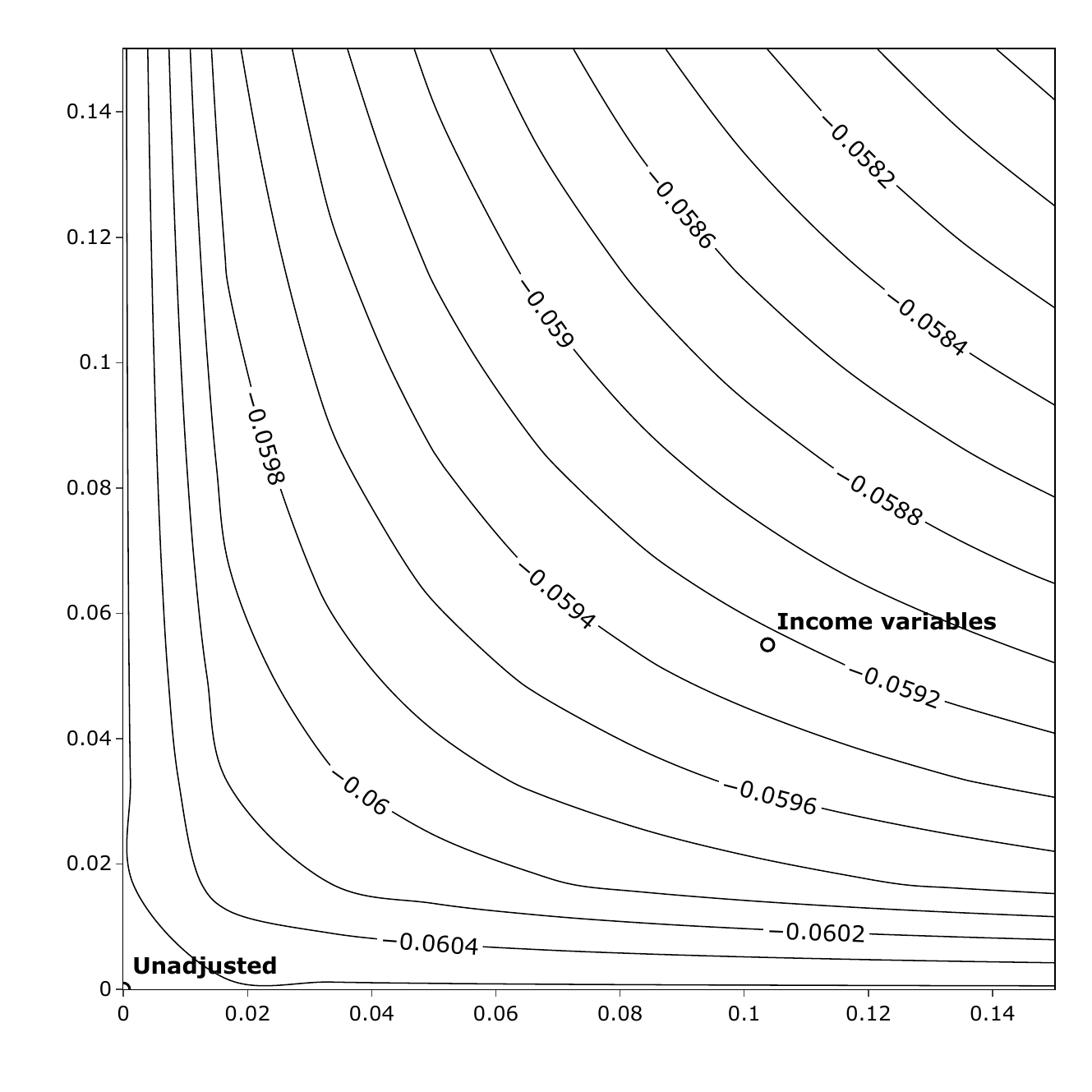}};

        \node[below=0pt of inc_theta_rhohat] (inc_theta_rho1)
        {\includegraphics[width=0.5\linewidth]{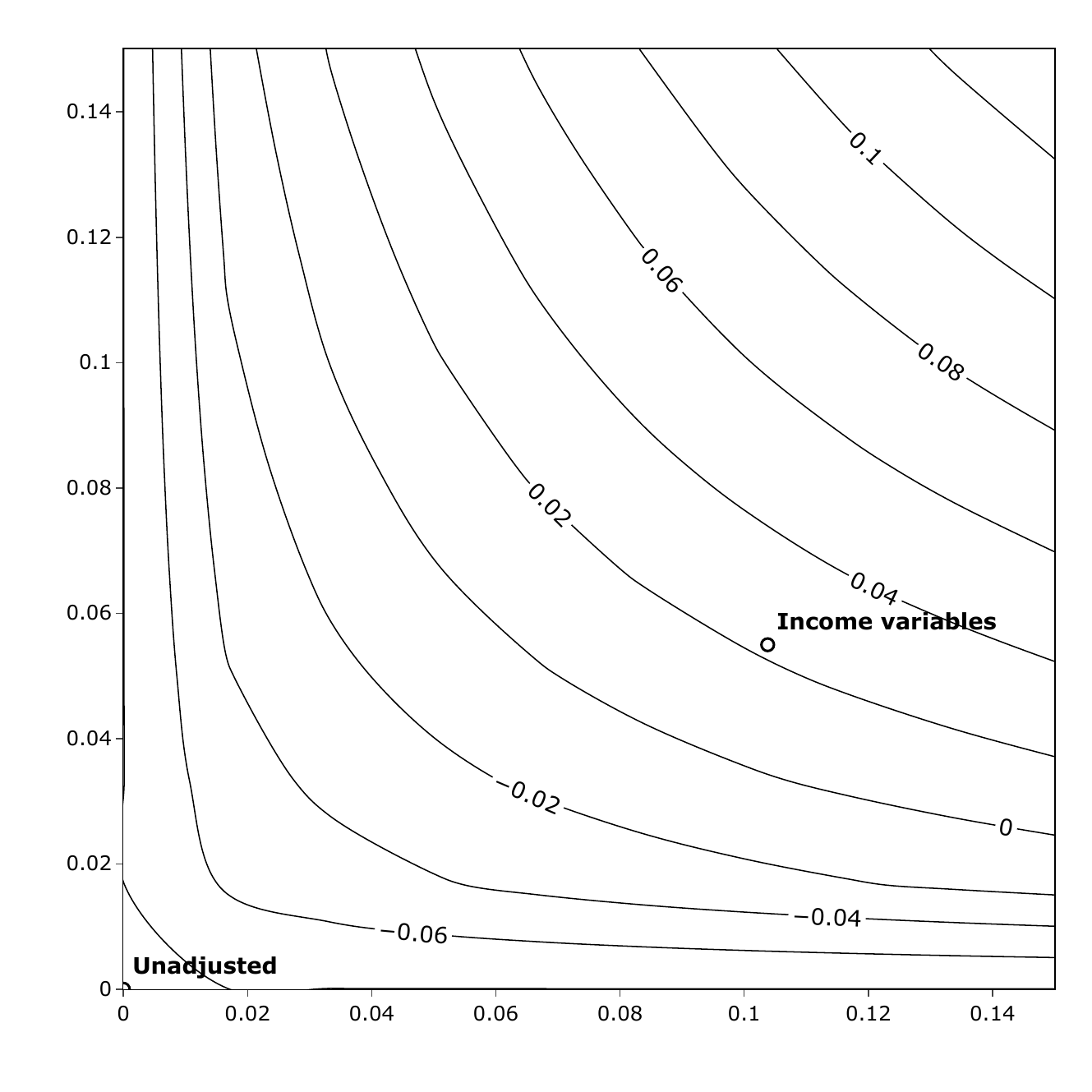}};
        \node[right=0pt of inc_theta_rho1] (inc_ci_rho1)
        {\includegraphics[width=0.5\linewidth]{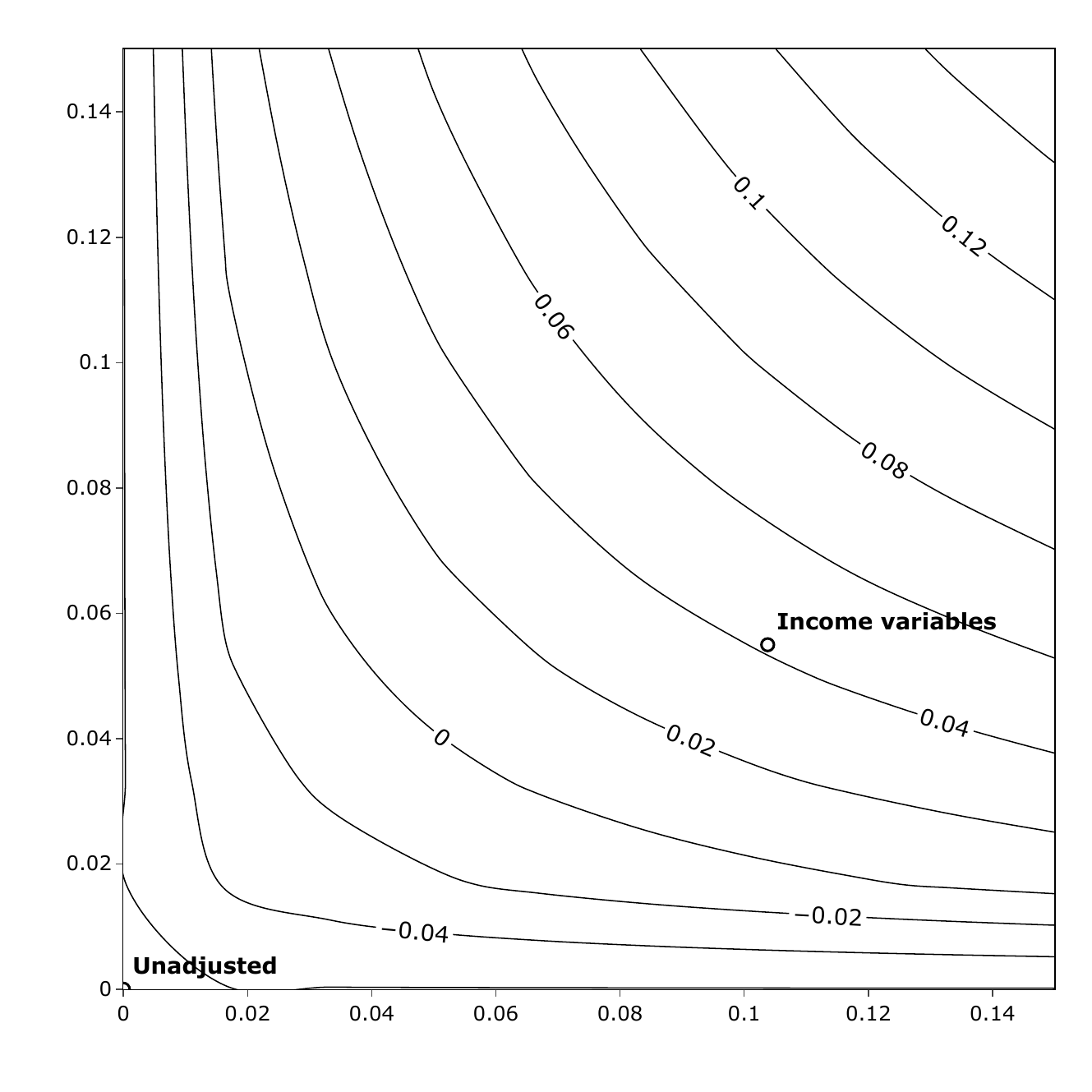}};

        \node[above=2mm of inc_theta_rhohat, font=\bfseries] {Bound on treatment effect};
        \node[above=2mm of inc_ci_rhohat, font=\bfseries] {Bound on CI};

        \node[left=8mm of inc_theta_rhohat, rotate=90, anchor=center, font=\bfseries, align=center]
        {As adversarial as \textit{all income vars.},\\$\bm{\rho = 0.014$}};
        \node[left=8mm of inc_theta_rho1, rotate=90, anchor=center, font=\bfseries, align=center]
        {Most adversarial case,\\$\bm{\rho = 1$}};

        \node[below=0mm of inc_theta_rho1, xshift=0.25\linewidth, font=\small]
        {Confounding strength w.r.t.\ treatment};
        \node[left=0mm of inc_theta_rho1, yshift=0.5\linewidth, rotate=90, font=\small]
        {Confounding strength w.r.t.\ outcome};

    \end{tikzpicture}
    \caption{Sensitivity contours using all income variables as the benchmarking confounder set.}
    \label{fig:sens_incset_2x2}
    \begin{minipage}{\textwidth}
        \footnotesize
        \emph{Notes:}
        The left column reports bounds on
        $\theta$ and the right column reports bounds on the upper limit of the 95\% CI.
        Axes give $(cf_d,cf_y)$, the treatment- and outcome-side confounding strengths.
        Rows vary the ``degree of adversity'', $\rho$.
        ``Unadjusted'' is $(cf_d,cf_y)=(0,0)$ and ``Income variables'' is the benchmark point implied by omitting all income covariates.
        See Section \ref{sec:sens_longstoryshort}.
    \end{minipage}
\end{figure}

The results are plotted in Figures \ref{fig:sens_privsub_2x2} and
\ref{fig:sens_incset_2x2}, respectively.
Each figure is a two-by-two matrix. The left-hand panels report bounds on the
treatment-effect estimate, while the right-hand panels report bounds on
the upper limit of the confidence interval (``Bound on CI''). Because
the baseline estimate of the effect of aid denial on incarceration is negative,
a negative value in the CI panels indicates that the effect remains statistically
different from zero even after allowing for unobserved confounding of the given
strength.\footnote{Intuitively, if the upper confidence limit is still
    below zero, the entire confidence interval remains negative.}
Within each panel, the horizontal axis ($cf_d$) represents confounding strength
with respect to the treatment channel, and the vertical axis ($cf_y$)
represents confounding strength with respect to the outcome channel.
The point labelled ``Unadjusted'' corresponds to the
main treatment-effect estimate under no unobserved confounding,
$(cf_d,cf_y)=(0,0)$, while the other labelled
point (``Private submission'' or ``Income variables'')
corresponds to the confounding strength implied by omitting that benchmark
covariate set. The contour lines show how the relevant bound changes as one
allows for progressively stronger confounding (moving up and left in the plot).
Finally, different rows assume different ``degrees of adversity'', $\rho$:
the upper row
sets $\rho$ equal to the value estimated from the benchmark-variable
omission exercise, while the lower row sets $\rho=1$,
the most adverse, worst-case, scenario for a given confounding strength.

Figure~\ref{fig:sens_privsub_2x2} indicates that benchmarking on
$private\_sub$ yields a confounder that is strong on the treatment
dimension ($cf_d$ is non-trivial) but very weak on the outcome
dimension ($cf_y$ is near zero). As a result, even if a confounder ``as strong
as private submission'' were entirely unobserved, the implied bounds remain
close to the unadjusted estimate. In particular, at the benchmark point the CI
upper bound remains below zero, implying that the estimated negative effect of
aid denial on incarceration is robust to this type of confounding. Intuitively,
a variable like private submission predicts well aid denial, but it
adds little explanatory power for incarceration once the rich set of
controls is included.
It should also be noted that, notwithstanding its estimated $\rho$ is very
adversarial (being close to one), its weak outcome-side strength means
that omitting it cannot generate large omitted-variable
bias\textemdash not even when $\rho = 1$.

By contrast, Figure~\ref{fig:sens_incset_2x2} considers the omission of the
entire set of income covariates, which implies substantially larger
outcome-side confounding strength ($cf_y$), while still displaying
similarly high treatment-side confounding.
However, the estimated degree of adversity is very low
($\hat\rho = 0.014$, close to zero).
As a result, an unobserved confounder comparable to the income covariates
would be unlikely to overturn either the sign or the statistical significance
of the main treatment-effect estimates. To flip the sign and render the
estimate statistically insignificant at the 5\% level,
one would need to assume an omitted confounder that is not only as strong as
the income covariates\textemdash conditional on the remaining
controls\textemdash but also maximally adversarial ($\rho = 1$).
This scenario appears implausible: even if unobserved confounding remains,
it is difficult to think of confounders with both
that magnitude and that degree of adversity.

\subsection{Whole-Sample Results}
The preferred specification might maximise internal validity\textemdash
minimising bias\textemdash at the expense of external validity,
hence leaving us questioning whether its associated estimates
only apply to a restricted and special subpopulation or
generalise to the whole population of legal aid applicants charged with a
serious crime.
Hence, I re-estimate the treatment effects on the whole sample, which includes
terminated applications, applications submitted by private lawyers,
defendants represented by panel lawyers, and those who
self-represent.
I find results largely consistent with the preferred estimates,
as shown in Table \ref{tab:tab:irm_one_ATT}, Appendix \ref{sec:full_sample}.

\section{Discussion \label{sec:discuss}}

Australian law protects the right to counsel, meaning that
those who are denied aid can typically afford a private lawyer\textemdash
this is precisely why they are denied aid.
As reported in Table \ref{tab:tab:irm_one_inhouse_noterm_norep},
on the one hand,
I find that such private lawyers provide their clients with
the benefit of lowering their chances of incarceration by about 10 p.p.
The high-stakes nature of such trials for serious crimes, where the
mean (positive) incarceration length is 46.4 months\textemdash almost 4 
years\textemdash makes this 10 p.p. effect particularly consequential.

On the other hand, first, the evidence of an effect on incarceration length is
not strong, looking across specifications, and not large (point estimates of
-0.5 and -1.8 months for ATT and ATE, respectively).
Second, once we condition on receiving a (positive) jail sentence, being
represented by privately-funded lawyers predicts a 5-6 months \textit{increase}
of the defendant's incarceration spell.
As mentioned in Section \ref{sec:disclaimer},
the results concerning incarceration length, conditional on incarceration,
differ from the main findings in terms of their identification 
credibility\textemdash they measure correlations whose causal interpretation is 
uncertain.
Despite this caveat, I included them because the sign of these
estimates\textemdash showing that private lawyers increase jail time conditional
on incarceration\textemdash is consistent with the broader picture outlined
below.

Indeed, this pattern of findings related to incarceration
can be explained in conjunction with the estimated negative
effect of being privately represented on the probability of a
guilty plea (-5 p.p.).
We know that private lawyers are less time-efficient at handling legal
aid cases than in-house lawyers \citep[see][]{ooi2019}\footnote{
    Legal aid cases handled by panel lawyers are ``more likely to be finalised
    at a later stage in proceedings and are more likely to be finalised in the
    Higher Court'' \citep{ooi2019}
}. We also know from the present dataset that
legal aid applications have never been refused on a lack-of-funding basis,
notwithstanding a finite budget.
Finally, we know from \cite{millane2023} that Legal Aid staff sustains a
heavy workload.
Taken together, these facts and estimates suggest the following.

First, in-house lawyers face strong incentives to handle their caseloads
efficiently, consistent with the program's goal of maximising access to
aid for all eligible applicants while being subject to a fixed budget. 
Second, to manage this volume, they
appear to rely more heavily (than private counterfactual lawyers) on
guilty pleas and associated negotiations. This is consistent with the observed
pattern, where shorter sentences are negotiated at the expense of more 
frequent incarcerations, as part of the plea agreement.

Moreover, a complementary factor in reducing incarceration length for those
given a jail sentence could be the free access
to expert reports that legal aid recipients enjoy. Indeed, while expert
reports (for example,  by a psychiatrist) can be expensive, particularly
for low-income defendants not qualifying for aid, legal aid recipients can
access them free of charge thanks to the presence of in-house experts.
A more frequent use of experts by in-house lawyers compared to privately-funded
lawyers may allow the former to highlight mitigating factors more effectively,
and thus lead to a more lenient jail sentence.

In summary, the estimated effects on other outcomes
show null effects on unconditional
incarceration but negative effects on plea bargaining,
consistent with public lawyers opting for plea bargaining in situations where
private lawyers prefer to run the case.
If both public and private lawyers succeed in their respective strategies,
plea bargaining would result in shorter incarceration spells at the expense of
accepting some jail time, while running the case reduces incarceration risk
but potentially lengthens the incarceration spell.
On average, these two strategies may offset each other,
resulting in a null net effect on unconditional incarceration duration,
as my estimates indicate.
I include these findings as excluding them would detract
from this important point, which finds support in the data.
A final remark\textemdash
it is not surprising that private representation causes an increase in fines,
as the judge would see it as a signal that the defendant is better
able to pay them.

\subsection{Channels}
While the above results invite some nuance when comparing the performance of
private and in-house lawyers, the stronger performance of private lawyers
at avoiding any incarceration time to their clients (9.7 p.p., or 17\%
compared to the sample mean) warrants further discussion.
There are two main mechanisms that could be driving this incarceration
gap: self-selection and workload. According to the self-selection channel,
the greater financial rewards of privately-funded work could be attracting
more experienced lawyers and barristers. According to
the workload channel, the heavy workload of Legal Aid (in-house) legal
practitioners creates an incentive to spend less time on their cases,
ultimately affecting incarceration chances.
Both channels are supported
by the qualitative and quantitative evidence gathered in a recent report
commissioned by National Legal Aid \citep{millane2023}. This highlights
both the increasing gap in remuneration
between privately- and publicly-funded cases and the heavy workload that
legal aid staff has been facing over the years.

While the dataset used in the present study does not allow me to disentangle
the two channels\footnote{
    In particular, the lack of data on Aboriginal and Torres Strait Islander
    cases and on the number of legal practitioners working for Legal Aid
    prevents me from measuring the workload of the latter and how it correlates
    with court outcomes. In other words, this lack of data prevents me from
    disentangling the workload channel from the self-selection channel.
}, they are both likely to be
driving the gap and are both rooted in a funding shortage.
Indeed, funding has declined by 3\% per capita in the last decade, and
the gap between the market rate and the rate paid by Legal Aid to private
lawyers has only widened \citep{millane2023}.
This goes against the recommendations of a federal
Productivity Commission, which identified this issue almost a decade ago
\citep{pc2014}

The pay gap issue extends to (in-house) public lawyers and barristers.
Indeed, data from the 2008/09 financial year show that
public and private lawyers earned, respectively, A\$86,700 and A\$94,800
on average \citep{forell2010},
while public and private barristers earned,
respectively, A\$106,184 \citep{forell2010} and A\$361,850
\citep{abs_legal_2008}\footnote{
    Estimate calculated from ABS data dividing the income generated by
    private barristers in the 2007-08 financial year (A\$1.4b)
    by their number in the same period (3,869 private barristers).
    Public barristers are to be intended as senior public lawyers acting as
    barristers.
}
on average.

As mentioned, the workload that needs to be managed by public lawyers
is also likely to be playing a role.
Anecdotal evidence from \cite{millane2023} and \cite{smh2011} supports the
view that such a heavy workload represents an issue.
This could explain why Legal Aid clients are 5 p.p. more likely to plead guilty,
as the pressure coming from a high volume of cases creates an incentive
to pursue time-efficient solutions, particularly for cases where
the probability of incarceration is high.

\section{Conclusion \label{sec:conclusion} }
Legal aid to indigent defendants is omnipresent in developed countries.
It is commonly justified on social justice grounds, but this
should not preclude us from investigating it quantitatively.
Yet, it has remained largely unstudied.
Measuring its impact is important from an accountability point of view
\textemdash
that is, to compare public and private defence\textemdash but also
to understand its broader role within the criminal justice system.

This paper makes the first moves in filling this gap,
studying the effect of denying legal aid on court outcomes.
Thanks to rich administrative data from New South Wales, Australia,
a double machine learning approach can credibly learn the unknown
treatment assignment function and identify treatment effects.
I find that, under the current system, denying aid does not negatively impact
the defendants' court outcomes. Conversely, it reduces their probability
of incarceration by about 8.1 p.p., with such defendants mostly hiring a private
lawyer.
This suggests that a performance gap exists between private and
public legal representation, further supported by our analysis comparing
publicly- and privately-funded representation.

While I highlighted some nuance to this story, the present findings do
point to a sizeable performance gap. At the root of the gap sits a
problem common in welfare program: funding is limited and policymakers
have to face an uncomfortable quantity-quality tradeoff.
Looking at the extreme cases for the purpose of clarity\textemdash
if policymakers concentrated all their available budget
on the poorest applicant,
they would leave many other indigent applicants without representation
(the problem of \textit{access}). Conversely, if they spread the budget across
the whole population, then per capita resources would be very limited.
This trade-off can only be resolved by political institutions, and
measuring it is a necessary first step to opening a constructive,
evidence-based, debate.

Beyond the magnitude and sign of these estimates,
such a DML approach to impact evaluation gives policy-makers a new tool to
study the effects of public programs, either substituting traditional
econometric methods where their assumption are not credible, or complementing
them.

\bibliographystyle{apa}
\bibliography{legal_aid_biblio}

\FloatBarrier
\clearpage
\appendix

\section{Variable Descriptions \label{appx:var_des}}

\begin{ThreePartTable}
    \begin{TableNotes}[para,flushleft]
        \footnotesize
        \item \textit{Notes:} (1): Gross assessable income encompasses all income sources that are subject to assessment.
        (2): A Financially Associated Person (FAP) allowance is a type of deduction applied during the assessment of an
        applicant's financial eligibility for legal aid. This allowance is deducted for each FAP whose income is included
        in the applicant's assessable income. The purpose of this allowance is to account for the financial
        responsibilities the applicant has towards these associated individuals, typically a spouse or partner, and
        hence provide a more accurate representation of their disposable income.
        Variables in upper case are sourced from the Legal Aid database, those in lower case from the ROD database.
    \end{TableNotes}
    \begin{longtable}{l p{10cm}}
        \caption{Table of Variables and Descriptions\label{tab:variables}}                                                    \\
        \toprule
        \textbf{Variable}                & \textbf{Description}                                                               \\
        \midrule
        \endfirsthead

        \toprule
        \textbf{Variable}                & \textbf{Description}                                                               \\
        \midrule
        \endhead

        \midrule
        \multicolumn{2}{r}{\small\itshape Continued on next page}                                                             \\
        \midrule
        \endfoot

        \bottomrule
        \insertTableNotes
        \endlastfoot

        dependants                       & Number of dependants                                                               \\
        child\_support\_dependants       & Number of dependants on child support                                              \\
        inc\_pension\_benf               & Gross (that is, pre-deductions) pension income from all types of pension payments  \\
        assess\_inc\_pension\_benf       & Gross pension income from assessable types of pension payments$^{1}$               \\
        assess\_inc\_from\_employment    & Gross assessable employment income                                                 \\
        assess\_inc\_busn\_self          & Gross assessable self-employment income                                            \\
        assess\_inc\_child\_support      & Gross assessable child-support income                                              \\
        assess\_inc\_oth                 & Other gross assessable income                                                      \\
        assess\_total\_inc               & Total gross assessable income                                                      \\
        exp\_tax                         & Total gross income minus deductions                                                \\
        exp\_rent                        & Tax expenses                                                                       \\
        exp\_mtgage                      & Rent expenses                                                                      \\
        exp\_board                       & Mortgage expenses                                                                  \\
        exp\_rates                       & Board expenses                                                                     \\
        exp\_child\_care                 & Costs associated with local government charges on property ("council rates")       \\
        exp\_child\_support              & Childcare expenses                                                                 \\
        assess\_exp\_tax                 & Child support expenses                                                             \\
        assess\_exp\_rent                & Assessable tax expenses                                                            \\
        assess\_exp\_mtgage              & Assessable rent expenses                                                           \\
        assess\_exp\_board               & Assessable mortgage expenses                                                       \\
        assess\_exp\_rates               & Assessable board expenses                                                          \\
        assess\_exp\_wkly\_housing       & Assessable council rates expenses                                                  \\
        assess\_exp\_child\_care         & Assessable weekly housing expenses                                                 \\
        assess\_exp\_child\_support      & Assessable childcare expenses                                                      \\
        assess\_exp\_fap\_depend         & Assessable child support expenses                                                  \\
        assess\_net\_inc                 & Assessable financially-associated person (FAP) expenses                            \\
        home                             & Primary home equity                                                                \\
        home\_mtgage                     & Outstanding primary home mortgage size                                             \\
        oth\_real\_estate                & Value of other real estate equity                                                  \\
        oth\_mtgage                      & Other outstanding mortgages size                                                   \\
        motor\_vehicle                   & Motor vehicle value                                                                \\
        motor\_vehicle\_loan             & Size of motor vehicle loan                                                         \\
        farm\_busn                       & Farm business equity                                                               \\
        farm\_busn\_mtgage               & Outstanding farm business mortgage                                                 \\
        cash                             & Value of cash assets                                                               \\
        oth\_assets                      & Assessable FAP allowance$^{2}$                                                     \\
        assess\_home                     & Value of other assets                                                              \\
        assess\_home\_mtgage             & Assessable value of other assets                                                   \\
        assess\_home\_excluded           & Total assessable assets related to FAP declared by the applicant                   \\
        assess\_net\_home                & Value of lump sum payments excluded from the Assets Test                           \\
        assess\_oth\_real\_estate        & Assessable net (that is, after accounting for liabilities and exclusions) assets   \\
        assess\_oth\_mtgage              & Assessable value of other real estate, excluding primary residence                 \\
        assess\_net\_oth\_real\_estate   & Cash assets excluded from the assessable cash total                                \\
        assess\_motor\_vehicle           & Other assets excluded from the assessable total                                    \\
        assess\_motor\_vehicle\_loan     & Net value of cash assets                                                           \\
        assess\_vehicle\_excluded        & Net value of other assessable assets                                               \\
        assess\_net\_motor\_vehicle      & Assessable value of home equity                                                    \\
        assess\_farm\_busn               & Assessable value of outstanding mortgage on the primary home                       \\
        assess\_farm\_busn\_mtgage       & Portion of home equity excluded from asset calculations                            \\
        assess\_farm\_busn\_excluded     & Net assessable home equity                                                         \\
        assess\_net\_farm\_busn          & Assessable value of other real estate holdings                                     \\
        assess\_cash                     & Outstanding mortgage on other real estate holdings                                 \\
        assess\_oth\_assets              & Net value of other real estate after liabilities                                   \\
        assess\_tot\_client\_fap\_assets & Assessable value of motor vehicles                                                 \\
        assess\_preclusion\_lsum\_asset  & Outstanding loan amount on motor vehicles                                          \\
        assess\_fap\_depend\_allowance   & Motor vehicle values excluded from asset calculations                              \\
        assess\_net\_assets              & Net assessable value of motor vehicles after deductions                            \\
        assess\_oth\_real\_estate\_excl  & Assessable value of farm business assets                                           \\
        assess\_cash\_excluded           & Outstanding mortgage on farm business assets                                       \\
        assess\_oth\_assets\_excluded    & Portion of farm business assets excluded from assessment                           \\
        assess\_net\_cash                & Net assessable value of farm business assets after liabilities                     \\
        assess\_net\_oth\_assets         & Assessable value of cash assets                                                    \\
        female                           & Indicator for female sex (0 if male)                                               \\
        submission\_year                 & Year in which the application was submitted                                        \\
        private\_sub                     & Indicator for whether the submission was made by private lawyer                    \\
        private\_ass                     & Indicator for whether a case was assigned to a private lawyer                      \\
        interpreter                      & Indicator for whether an interpreter was required                                  \\
        age                              & Age of the defendant                                                               \\
        disability                       & Indicator for the presence of a disability                                         \\
        domestic\_violence               & Indicates whether an individual's recorded offense is related to domestic violence \\
        \bottomrule
    \end{longtable}
\end{ThreePartTable}

\section{Other Eligibility Tests \label{appx:tests}}
The Jurisdiction test checks whether the applicant's matter is in a jurisdiction
or area of law covered by Legal Aid. Focusing on
Criminal Law, aid is available for most matters: Local Court,
District Court,
Supreme Court,
Court of Criminal Appeal and High Court,
Children's and Prisoner's matters. Aid is not available for cases where the
applicant has been caught in the act of committing violent crimes, for Local
Court hearings where there are no prospects of success, for cases related to
monies and/or property used or intended to be used to commit a crime
\citep[``tainted'', see][]{crimeAct} to start trials in the Local Court,
except for Apprehended Domestic Violence Orders trials.

The Merit test assesses what are the net benefits of providing or denying legal
aid to the applicant, and whether the applicant has a reasonable chance to
win the case. In practice, this test is used to deny aid for matters that
are not serious and/or have no chance of success. Serious criminal matters
automatically pass the Merit test. For other cases exempt from the
Merit test, refer to \citep[see]{laPolicies}.

\newpage
\section{Other Sensitivity Analyses \label{appx:other_sens}}

\begin{ThreePartTable}
  \begin{TableNotes}[para,flushleft]
    \footnotesize
    \item \textit{Notes: }
    The table reports standardized mean differences between treatment and
    control groups, unadjusted and adjusted using propensity score weights.
    The propensity scores (and hence the weights) are those used in the
    preferred specification and are estimated via random forests on the
    (within-transformed) main sample.
    The table is produced in \texttt{R} using \cite{cobalt}.

  \end{TableNotes}
  \begin{longtable}{l l r r}
    \caption{\label{tab:ps_bal_tab_main}Balance Table Before and After Propensity-Score Weighing}                    \\
    \toprule
    \textbf{Variable}                & \textbf{Type} & \textbf{Difference Unadjusted} & \textbf{Difference Adjusted} \\
    \midrule
    \endfirsthead

    \toprule
    \textbf{Variable}                & \textbf{Type} & \textbf{Difference Unadjusted} & \textbf{Difference Adjusted} \\
    \endhead

    \midrule
    \multicolumn{4}{r}{\small\itshape Continued on next page}                                                        \\
    \midrule
    \endfoot

    \bottomrule
    \insertTableNotes
    \endlastfoot

    dependants                       & Contin.       & 0.17                           & 0.07                         \\
    child\_support\_dependants       & Contin.       & 0.02                           & -0.01                        \\
    inc\_pension\_benf               & Contin.       & 0.02                           & 0.00                         \\
    assess\_inc\_pension\_benf       & Contin.       & 0.03                           & 0.00                         \\
    assess\_inc\_from\_employment    & Contin.       & 0.16                           & 0.11                         \\
    assess\_inc\_busn\_self          & Contin.       & 0.12                           & 0.10                         \\
    assess\_inc\_child\_support      & Contin.       & 0.03                           & 0.02                         \\
    assess\_inc\_oth                 & Contin.       & 0.10                           & 0.09                         \\
    assess\_total\_inc               & Contin.       & 0.18                           & 0.11                         \\
    exp\_tax                         & Contin.       & 0.13                           & 0.09                         \\
    exp\_rent                        & Contin.       & 0.23                           & 0.09                         \\
    exp\_mtgage                      & Contin.       & 0.03                           & 0.03                         \\
    exp\_board                       & Contin.       & 0.07                           & 0.03                         \\
    exp\_rates                       & Contin.       & 0.11                           & 0.04                         \\
    exp\_child\_care                 & Contin.       & 0.07                           & 0.05                         \\
    exp\_child\_support              & Contin.       & 0.10                           & 0.06                         \\
    assess\_exp\_tax                 & Contin.       & 0.14                           & 0.08                         \\
    assess\_exp\_rent                & Contin.       & 0.24                           & 0.08                         \\
    assess\_exp\_mtgage              & Contin.       & 0.04                           & 0.03                         \\
    assess\_exp\_board               & Contin.       & 0.07                           & 0.02                         \\
    assess\_exp\_rates               & Contin.       & 0.12                           & 0.04                         \\
    assess\_exp\_wkly\_housing       & Contin.       & 0.33                           & 0.14                         \\
    assess\_exp\_child\_care         & Contin.       & 0.07                           & 0.03                         \\
    assess\_exp\_child\_support      & Contin.       & 0.09                           & 0.05                         \\
    assess\_exp\_fap\_depend         & Contin.       & 0.23                           & 0.10                         \\
    assess\_net\_inc                 & Contin.       & 0.14                           & 0.10                         \\
    home                             & Contin.       & 0.21                           & 0.14                         \\
    home\_mtgage                     & Contin.       & 0.19                           & 0.14                         \\
    oth\_real\_estate                & Contin.       & 0.12                           & 0.12                         \\
    oth\_mtgage                      & Contin.       & 0.12                           & 0.12                         \\
    motor\_vehicle                   & Contin.       & 0.20                           & 0.11                         \\
    motor\_vehicle\_loan             & Contin.       & 0.12                           & 0.07                         \\
    farm\_busn                       & Contin.       & 0.04                           & 0.02                         \\
    farm\_busn\_mtgage               & Contin.       & 0.01                           & 0.01                         \\
    cash                             & Contin.       & 0.13                           & 0.08                         \\
    oth\_assets                      & Contin.       & 0.06                           & 0.06                         \\
    assess\_home                     & Contin.       & 0.22                           & 0.14                         \\
    assess\_home\_mtgage             & Contin.       & 0.12                           & 0.09                         \\
    assess\_home\_excluded           & Binary        & 0.00                           & 0.00                         \\
    assess\_net\_home                & Contin.       & 0.07                           & 0.05                         \\
    assess\_oth\_real\_estate        & Contin.       & 0.13                           & 0.12                         \\
    assess\_oth\_mtgage              & Contin.       & 0.13                           & 0.13                         \\
    assess\_net\_oth\_real\_estate   & Contin.       & 0.08                           & 0.08                         \\
    assess\_motor\_vehicle           & Contin.       & 0.21                           & 0.12                         \\
    assess\_motor\_vehicle\_loan     & Contin.       & 0.13                           & 0.07                         \\
    assess\_vehicle\_excluded        & Contin.       & -0.02                          & -0.03                        \\
    assess\_net\_motor\_vehicle      & Contin.       & 0.08                           & 0.06                         \\
    assess\_farm\_busn               & Contin.       & 0.05                           & 0.02                         \\
    assess\_farm\_busn\_mtgage       & Contin.       & 0.02                           & 0.01                         \\
    assess\_farm\_busn\_excluded     & Binary        & 0.00                           & 0.00                         \\
    assess\_net\_farm\_busn          & Contin.       & 0.04                           & 0.04                         \\
    assess\_cash                     & Contin.       & 0.14                           & 0.08                         \\
    assess\_oth\_assets              & Contin.       & 0.07                           & 0.06                         \\
    assess\_tot\_client\_fap\_assets & Contin.       & 0.13                           & 0.10                         \\
    assess\_preclusion\_lsum\_asset  & Binary        & 0.00                           & 0.00                         \\
    assess\_fap\_depend\_allowance   & Contin.       & 0.13                           & 0.05                         \\
    assess\_net\_assets              & Contin.       & 0.13                           & 0.10                         \\
    assess\_oth\_real\_estate\_excl  & Binary        & 0.00                           & 0.00                         \\
    assess\_cash\_excluded           & Binary        & 0.00                           & 0.00                         \\
    assess\_oth\_assets\_excluded    & Binary        & 0.00                           & 0.00                         \\
    assess\_net\_cash                & Contin.       & 0.14                           & 0.08                         \\
    assess\_net\_oth\_assets         & Contin.       & 0.07                           & 0.06                         \\
    female                           & Contin.       & 0.02                           & 0.01                         \\
    submission\_year                 & Contin.       & 0.08                           & 0.05                         \\
    private\_sub                     & Contin.       & 0.77                           & 0.48                         \\
    private\_ass                     & Binary        & 0.00                           & 0.00                         \\
    interpreter                      & Contin.       & 0.03                           & 0.02                         \\
    age                              & Contin.       & 0.06                           & -0.01                        \\
    disability                       & Contin.       & -0.13                          & -0.12                        \\
    domestic\_violence               & Contin.       & 0.03                           & 0.01                         \\
    \bottomrule
  \end{longtable}
\end{ThreePartTable}

\newpage

\begin{table}[!h]
\centering
\caption{\label{tab:tab:prv_sub_main}Share of Private Submissions Across Treatment Groups}
\centering
\begin{tabular}[t]{rrr}
\toprule
\multicolumn{1}{c}{ } & \multicolumn{2}{c}{Private Submission} \\
\cmidrule(l{3pt}r{3pt}){2-3}
Denied Aid & 0 & 1\\
\midrule
0 & 0.91 & 0.09\\
1 & 0.56 & 0.44\\
\bottomrule
\end{tabular}
\end{table}

\begin{table}[!h]
\centering
\caption{\label{tab:tab:pos_ASS_NET_ASSETS_main}Share of Positive Ass. Net Assets Across Treatment Groups}
\centering
\begin{tabular}[t]{rrr}
\toprule
\multicolumn{1}{c}{ } & \multicolumn{2}{c}{Any Ass. Net Assets} \\
\cmidrule(l{3pt}r{3pt}){2-3}
Denied Aid & 0 & 1\\
\midrule
0 & 0.99 & 0.01\\
1 & 0.91 & 0.09\\
\bottomrule
\end{tabular}
\end{table}

\begin{table}[!h]
\centering
\caption{\label{tab:tab:assets_stats}Summary Statistics for Ass. Net Assets by Treatment Group}
\centering
\begin{tabular}[t]{lrr}
\toprule
\multicolumn{1}{c}{ } & \multicolumn{1}{c}{Granted} & \multicolumn{1}{c}{Denied} \\
\cmidrule(l{3pt}r{3pt}){2-2} \cmidrule(l{3pt}r{3pt}){3-3}
Statistic & Value & Value\\
\midrule
Min. & 0.00 & 0.00\\
1st Qu. & 0.00 & 0.00\\
Median & 0.00 & 0.00\\
Mean & 325.15 & 15088.37\\
3rd Qu. & 0.00 & 0.00\\
\addlinespace
Max. & 974700.00 & 6476262.00\\
\bottomrule
\end{tabular}
\end{table}

\begin{table}
\centering
\caption{\label{tab:tab:irm_one_ATT_600_main}DML: ATT of Not Receiving Aid on Court Outcomes 
                   when income and assets are 600A\$ p.w. or less 
                   and 1,000A\$ or less, respectively.}
\centering
\begin{threeparttable}
\begin{tabular}[t]{lllllll}
\toprule
  & Coef. & s.e. & l-CI & r-CI & p-value & Obs.\\
\midrule
Reduced charges & -0.018 & 0.010 & -0.038 & 0.002 & 0.077 & 16007\\
Reduced seriousness & -0.006 & 0.007 & -0.019 & 0.007 & 0.391 & 14206\\
Incarcerated - extensive & -0.075 & 0.009 & -0.094 & -0.057 & 0.000 & 16005\\
Incarceration (mth.) & 0.286 & 0.768 & -1.219 & 1.791 & 0.710 & 16007\\
Incarceration - intensive (mth.) & 5.677 & 1.448 & 2.839 & 8.514 & 0.000 & 8520\\
Guilty plea to highest charge & -0.050 & 0.010 & -0.069 & -0.031 & 0.000 & 16007\\
Fined & 0.052 & 0.007 & 0.039 & 0.065 & 0.000 & 15999\\
\bottomrule
\end{tabular}
\begin{tablenotes}
\item \textit{Notes:} The table presents Double Machine Learning ATT 
  estimates for the effect of being denied aid on case-level court outcomes. 
  The IRM model is applied to cases where the defendant earns a 
  net assessable income of 600A\$ per week or less and owns net assessable 
  assets valued at 1000A\$ or less. 
  The chosen ML method is Random Forests.
\end{tablenotes}
\end{threeparttable}
\end{table}

\begin{table}
\centering
\caption{\label{tab:tab:irm_one_ATT_inhouse_noterm_nocovid}DML: ATT of Not Receiving Aid on Court Outcomes
                   in a sample without terminated grants, panel lawyers, nor  
                   submission years above 2019}
\centering
\begin{threeparttable}
\begin{tabular}[t]{lrrrrrr}
\toprule
  & Coef. & s.e. & l-CI & r-CI & p-value & Obs.\\
\midrule
Reduced charges & -0.012 & 0.014 & -0.040 & 0.017 & 0.4194 & 13500\\
Reduced seriousness & -0.007 & 0.006 & -0.020 & 0.005 & 0.2486 & 11965\\
Incarcerated - extensive & -0.093 & 0.011 & -0.116 & -0.071 & 0.0000 & 13498\\
Incarceration (mth.) & 0.386 & 0.794 & -1.170 & 1.942 & 0.6269 & 13500\\
Incarceration - intensive (mth.) & 6.965 & 1.474 & 4.076 & 9.854 & 0.0000 & 6991\\
Guilty plea to highest charge & -0.051 & 0.012 & -0.074 & -0.029 & 0.0000 & 13500\\
Fined & 0.057 & 0.010 & 0.038 & 0.075 & 0.0000 & 13492\\
\bottomrule
\end{tabular}
\begin{tablenotes}
\item \textit{Notes:} The table presents Double Machine Learning ATE
  estimates for the effect of not receiving aid (having aid denied or
  terminated) on case-level court outcomes.
  The IRM model is applied to all cases where aid was not terminated, 
  where the defendant was not represented by a panel lawyer, and the case was 
  not submitted in years after 2019. 
  While flexibly including the submission-year covariate should account for 
  time-trends, this robustness checks serve to reduce concerns that 
  the COVID-19 pandemic years are not driving the main findings. 
  The chosen ML method is Random Forests.
 Finally, court-level fixed effects are included via a within transformation.
\end{tablenotes}
\end{threeparttable}
\end{table}

\begin{table}
\centering
\caption{\label{tab:tab:irm_one_ATT_fail_inc_assets}DML: ATT of Being Denied Aid Aid on Court Outcomes
                   in a restricted sample}
\centering
\begin{threeparttable}
\begin{tabular}[t]{lrrrrrr}
\toprule
  & Coef. & s.e. & l-CI & r-CI & p-value & Obs.\\
\midrule
Reduced charges & 0.005 & 0.036 & -0.065 & 0.076 & 0.8835 & 1192\\
Reduced seriousness & -0.005 & 0.022 & -0.049 & 0.038 & 0.8086 & 939\\
Incarcerated - extensive & -0.100 & 0.034 & -0.166 & -0.034 & 0.0029 & 1192\\
Incarceration (mth.) & -3.091 & 2.403 & -7.801 & 1.620 & 0.1984 & 1192\\
Incarceration - intensive (mth.) & 4.913 & 4.956 & -4.800 & 14.626 & 0.3215 & 308\\
Guilty plea to highest charge & 0.016 & 0.034 & -0.050 & 0.081 & 0.6389 & 1192\\
Fined & 0.072 & 0.021 & 0.031 & 0.113 & 0.0006 & 1192\\
\bottomrule
\end{tabular}
\begin{tablenotes}
\item \textit{Notes:} The table presents Double Machine Learning ATT
  estimates for the effect of having aid denied terminated on case-level 
  court outcomes.
  The IRM model is applied to cases where: (i) aid was either granted or denied 
  (not terminated), (ii) the defendant was represented by in-house lawyer, and 
  (iii) the defendant failed either the income or the assets test.
  The chosen ML method is Random Forests.
\end{tablenotes}
\end{threeparttable}
\end{table}

\newpage

%

\begin{table}[!h]

    \caption{\label{tab:aipw} AIPW: ATT of Denying Aid on Court Outcomes}
    \centering
    \begin{threeparttable}
        \begin{tabular}[t]{lrrrrrr}
            \toprule
            Outcome                          & Coef.  & s.e.  & l-CI   & r-CI   & p-value & Obs.  \\
            \midrule
            Reduced charges                  & -0.034 & 0.010 & -0.053 & -0.015 & 0.0003  & 16854 \\
            Reduced seriousness              & -0.014 & 0.007 & -0.027 & 0.000  & 0.0524  & 14875 \\
            Incarcerated - extensive         & -0.072 & 0.010 & -0.092 & -0.053 & 0.0000  & 16852 \\
            Incarceration (mth.)             & 5.072  & 1.532 & 2.070  & 8.074  & 0.0009  & 16854 \\
            Incarceration - intensive (mth.) & 8.209  & 1.567 & 5.138  & 11.281 & 0.0000  & 8753  \\
            \addlinespace
            Guilty plea to highest charge    & -0.025 & 0.009 & -0.042 & -0.007 & 0.0071  & 16854 \\
            Fined                            & 0.054  & 0.007 & 0.040  & 0.068  & 0.0000  & 16846 \\
            \bottomrule
        \end{tabular}
        \begin{tablenotes}
            \item \textit{Notes:} Augmented inverse probability weighting with propensity score estimated via logistic regression and outcome regression via GLM (OLS for continuous outcomes and logit for binary), using covariates: ASS\_NET\_INC, ASS\_NET\_ASSETS, DEPENDANTS, ASS\_NET\_CASH, ASS\_TOT\_CLIENT\_FAP\_ASSETS, ASS\_FAP\_DEPEND\_ALLOWANCE, private\_sub.
        \end{tablenotes}
    \end{threeparttable}
\end{table}

\begin{figure}[H]
    \centering

    \subfloat[Centred data\label{fig:top}]{
        \includegraphics[width=4.5in]{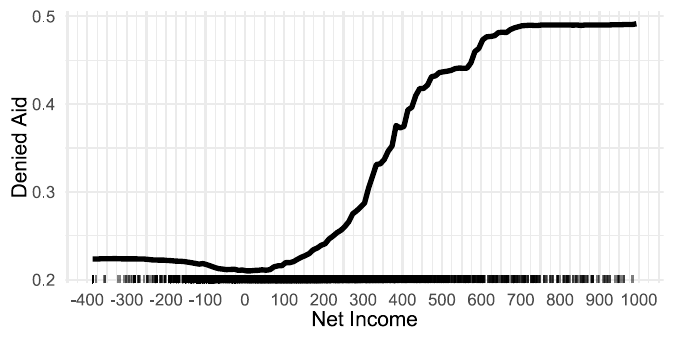}
    }

    \vspace{0.6em}

    \subfloat[Uncentred data\label{fig:bottom}]{
        \includegraphics[width=4.5in]{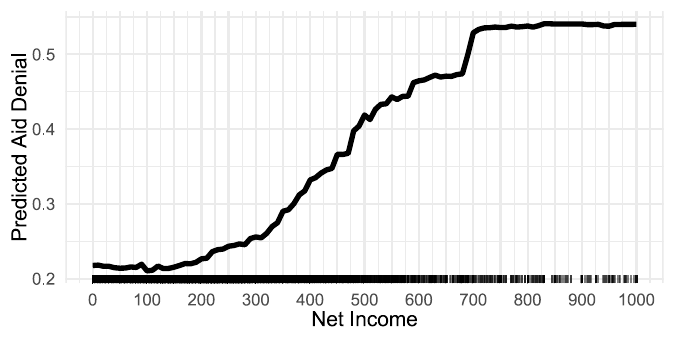}
    }

    \caption{Predicted Probability of Aid Denial}
    \label{fig:pdp_inc}
    \begin{minipage}{\textwidth}
        \footnotesize
        \emph{Notes:} This PDPs
        illustrate how the model's predicted probability of having a legal aid
        application denied changes as \emph{Assessable Net Assets}
        varies. Panel (A) uses court-demeaned net income,
        so the horizontal axis is interpreted as deviations from a court's
        mean net income. Panel (B) uses
        uncentred net income in levels\textemdash this is easier to interpret
        but is produced from a version of the main model without fixed effects.
        The black marks on the x-axis (rug plot)
        indicate the empirical
        distribution of net income in the estimation sample.
    \end{minipage}
\end{figure}

\section{DML Estimates Results on Whole Sample}
In this section, I provide evidence that the key main results
generalise to the full population of NSW Legal Aid applicants
charged with serious (indictable) crimes.

I start by estimating the DML model on the whole sample
(Table \ref{tab:tab:irm_one_ATT}).
In other words, I include applicants to which aid was initially granted but then
terminated (henceforth, \textit{terminated cases}), and publicly-funded cases
(that is, that have been granted legal aid) run by private lawyers.
This also implies increasing the sample size (lowering variance) at the expense
of some additional bias.
Then, I test the robustness of these
results and refine them.

\subsection{Full-Sample Estimates \label{sec:full_sample}}
Table \ref{tab:tab:irm_one_ATT} reports the ATT estimates
for the full-sample specification.
This specification includes, in the control group, cases for which legal aid
was granted and where the defendant was represented either by an in-house
or panel lawyer\textemdash that is, either by a Legal Aid lawyer, or by a
publicly-funded private lawyer. In the treatment group, it includes
cases for which aid was either denied or terminated, and where the
defendant was either represented by a private lawyer or self-represented.
These estimates are more likely to be biased (or to be more biased) than the
main estimates because (i) panel lawyers introduce an element of selection,
as they can choose which cases to run, and (ii) terminated cases introduce
a source of heterogeneity in the treatment group, as these applications
might have been terminated because of the applicant's misconduct.

Starting with the two most important outcomes, I find that denying or
terminating aid decreases the chances of incarceration by 11.5 p.p. and
the time in jail by 4.2 months. As I show below, the sign of these
estimates is robust across multiple specifications. Other robust findings are
that denying or terminating aid reduces the chances of pleading guilty
by 3.6 p.p. and increase the probability of being fined by 8.3 p.p.
These estimates make sense given most treated defendants are privately
represented and are evidence that denying aid does not harm legal aid
applicants under the current rules.

Table \ref{tab:tab:irm_one_ATT_inhouse_noterm} also shows that denying or
terminating aid: reduces the chances of some charges being dropped of withdraw;
leads to an increase in the share of charges for which the defendant is
found guilty; and reduces the defendant's chances to have their most serious
charge dropped or withdrawn. However, I show below that these findings are not
robust.

\begin{table}
\centering
\caption{\label{tab:tab:irm_one_ATT}DML: ATT of denying aid on Court Outcomes 
                   (whole sample)}
\centering
\begin{threeparttable}
\begin{tabular}[t]{lrrrrrr}
\toprule
  & Coef. & s.e. & l-CI & r-CI & p-value & Obs.\\
\midrule
Reduced charges & -0.034 & 0.008 & -0.051 & -0.018 & 0.0000 & 34668\\
Reduced seriousness & -0.012 & 0.005 & -0.022 & -0.002 & 0.0198 & 30172\\
Incarcerated - extensive & -0.115 & 0.008 & -0.130 & -0.100 & 0.0000 & 34665\\
Incarceration (mth.) & -4.105 & 0.677 & -5.432 & -2.779 & 0.0000 & 34668\\
Incarceration - intensive (mth.) & 3.030 & 1.392 & 0.302 & 5.758 & 0.0295 & 18844\\
Guilty plea to highest charge & -0.036 & 0.008 & -0.051 & -0.020 & 0.0000 & 34668\\
Fined & 0.084 & 0.006 & 0.072 & 0.095 & 0.0000 & 34655\\
\bottomrule
\end{tabular}
\begin{tablenotes}
\item \textit{Notes:} The table presents Double Machine Learning ATT 
  estimates for the effect of denying aid or terminating aid 
  on court outcomes. The IRM model is applied to the 
  whole sample. The chosen ML method is Random Forests.
\end{tablenotes}
\end{threeparttable}
\end{table}

\begin{figure}[!hp]
    \includegraphics[width=\linewidth]{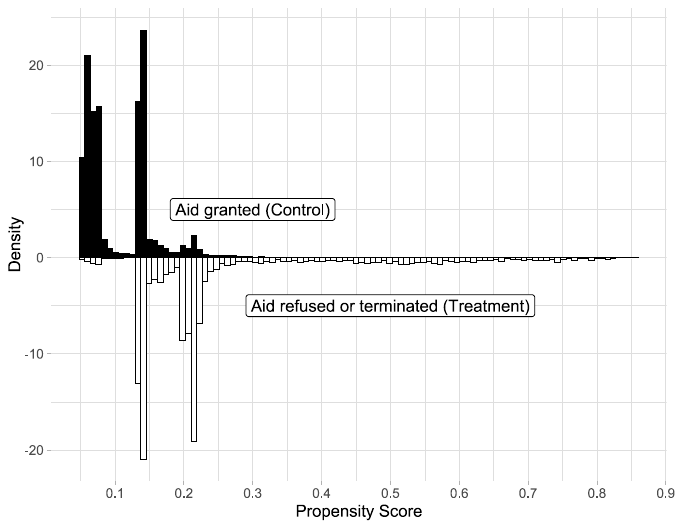}
    \caption{Propensity score}
    \label{fig:ps}
\end{figure}

To explore the sensitivity of these estimates, I plot their associated
propensity score by treatment group in Figure \ref{fig:ps}\footnote{
    The propensity score densities appear non-normally distributed. This
    does not represent an issue for the estimation of treatment effects.
    Their distribution could be due to a number of factors, including
    the highly zero-inflated distribution of key treatment determinants such as
    income and assets.
}.
Indeed, \cite{chern2018}'s Interactive Regression Model involves the
estimation of a propensity score\textemdash here via random forests.
We can visually inspect it\textemdash I break it down by treatment
group\textemdash to check that the common support assumption is reasonable
and to explore whether there are areas where the two propensity score
densities do not overlap.
As Figure \ref{fig:ps} shows, no propensity score is equal to one or zero,
providing evidence in favour of the common support assumption.
However, the high density around [0.05, 075] interval warrants caution,
as the estimator assigns high weight to these observations.
Moreover, while there is a good degree of overlap between the propensity
scores of the treated and control groups,
the treated group's density has a long right tail
where it has little to no overlap with the corresponding density for the control
group. While this is simply reflecting the imbalance between treatment groups,
it is important to investigate how sensitive the results are to the exclusion
of this tail.

\section{Fuzzy Regression Discontinuity \label{appx:frd}}
In this section, I attempt a fuzzy regression discontinuity analysis of the
local average treatment effect of being denied aid on charge-level outcomes.

\subsection{Empirical Strategy \label{sec:rdd}}
As mentioned above, some applicants who do not pass the means test, earning
$\$400$ or more per week in income,
    but receive legal aid anyway by \textit{discretion}.
    This implies that the threshold rule is probabilistic rather than deterministic,
    and hence that a fuzzy RD approach is warranted.
    I attempt a fuzzy regression discontinuity (FRD) design to estimate
    the effect of being represented by a private rather than public lawyer
    on the marginal applicant for the population
    at the cutoff, $c$.
In this FRD, the running variable is formal income and the cutoff value is
400\$ p.w.

\subsubsection{Identification}
The identification of local average treatment effects in fuzzy RDDs requires
three main assumptions: continuity, monotonicity and exogeneity.

The continuity assumption states that the conditional expectation function (CEF)
of the potential outcomes is continuous at the
cutoff. Formally, it states that
$\mathbb{E}\left[Y_i(D_i) \mid X_i=x\right]$ is continuous in $X$ at $c$.
This assumption can be invalidated by manipulations of the
running variable
\citep[with the exception of noisy manipulation, as shown in][]{ishihara2022}.
In the present paper's context, manipulation would occur if either the lawyer
processing the application or the applicant were able to systematically
misreport the true formal income (running variable), for instance
underreporting it to give aid to those not passing the means test.
The analyst would not be able to observe the true scores and determine which
were manipulated. Manipulation is unlikely to be an issue here, as formal
income is hard to doctor. While some applicants do not pass the means test
and still receive aid, I can identify them. It is not due to misreporting
of their earning, it is due to discretion\textemdash resulting in a fuzzy
design.

I test the continuity assumption primarily relying on the \cite{mccrary08} test
of discontinuity in the density function of the running variable around the
cutoff. Under the auxiliary assumption of one-sided
manipulation, this is the only valid test of the
continuity assumption \citep{ishihara2022}. The one-sided manipulation
assumption states that all noise-free (or \textit{precise}) manipulation
must be guided by either an incentive to treat or untreat but not both
simultaneously\textemdash as can happen when manipulators differ in their
motivations.

Another key assumption is the monotonicity assumption.
Adapted to the present context, where I define treatment as not getting aid,
the monotonicity assumption states that, for any
defendant, earning marginally more than 400\$ p.w. income cutoff does not make
them less likely to hire a private lawyer.

The exogeneity assumption, typical of IV estimation, is local in RDDs.
It implies that, in the neighbourhood of the cutoff,
not being eligible for legal aid (the instrument)
affects court outcomes only via its effect on the
probability of hiring a private lawyer (the treatment)\textemdash the exclusion
restriction assumption.
It is hard to imagine how a marginal difference in the
weekly income of an applicant
would affect anything other than the probability of receiving aid, or,
equivalently, of hiring a private lawyer.
This assumption also implies assuming no manipulation, that is, that a
defendant's formal income
(the running variable) does not depend on any potential
knowledge of the treatment effects.

\subsubsection{Treatment and Treatment Group}
For consistency with the RDD literature, I define treatment as being
on the right side of the cutoff. In this FRD application, it implies that (i)
the ``theoretical'' treatment assignment $T_i$ is ``having a net
assessable income of more than 400 AUD p.w.'', and (ii)
actual treatment $D_i$ is ``being represented by a private lawyer''.
In my sample, defendants who hire a private lawyer were refused aid.
This means that they applied for aid, but it was either not granted or it was
initially granted but later terminated before the end of the proceedings.

\subsection{Density Test}
\begin{figure}[h]
    \centering
    \includegraphics[width=\linewidth]{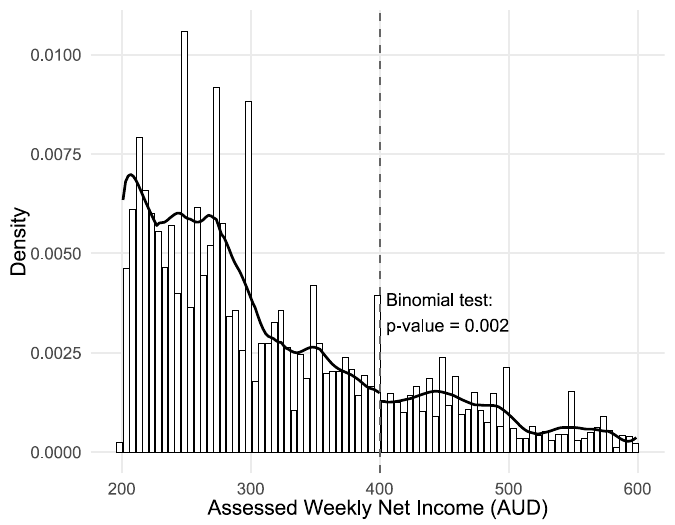}
    \caption{Density Tests}
    \label{fig:densities}
\end{figure}

\begin{figure}
    \includegraphics[width=\linewidth]{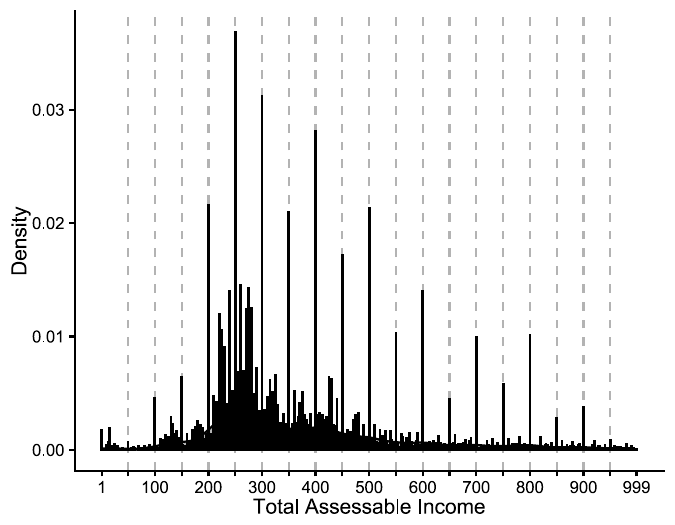}
    \caption{Histogram and density of assessable income}
    \label{fig:hist_dens_inc}
\end{figure}

Preliminary to any regression discontinuity analysis
is a test for the continuity of the density of the
running variable\textemdash here, net assessable income.
As visualised in Figure \ref{fig:densities}, there is an abnormal
concentration of defendants earning 400 AUD per week in net assessable income.
While, on the one hand, Figure \ref{fig:hist_dens_inc}
shows that this phenomenon affects
net assessable income at each 50\$ mark, on the other hand, it does seem more
pronounced at 400. Moreover, we don't know if the rounding is one-sided or,
as would be ideal, two-sided\textemdash with both values lower and higher than
400 being rounded to 400. Finally, we do not know if the rounding behaviour is
endogenous to the provision of aid.

Given this strong violation of the continuity assumption, I decide against
an RDD approach and in favour of a double machine learning one.
The only way forward from this violation would be taking a donut
hole approach \citep{barreca2011}, where the continuity assumption is restored
by dropping observations around the cutoff and thus imposing stronger
assumptions on the data generating process.
Moreover, the number of observations around the cutoff are already low in the
original sample\textemdash raising sample size concerns and limiting our ability
to capture small- and medium-sized effect\textemdash and the donut approach
would only make this issue more severe.

%
%
%
%
%
%
%
%
%


\end{document}